\documentclass[11pt,a4paper]{article}
\pdfoutput=1

\usepackage[affil-it]{authblk}
\usepackage[a4paper,textwidth=16.5cm,textheight=24cm,centering]{geometry}

\usepackage[]{amsmath}
	\numberwithin{equation}{section}
\usepackage[]{amssymb}
\usepackage{amscd}
\usepackage{amsfonts}
\usepackage{mathrsfs}

\DeclareMathOperator{\R}{\mathbb{R}}
\DeclareMathOperator{\C}{\mathbb{C}}

\DeclareMathOperator{\Z}{\mathbb{Z}}
\DeclareMathOperator{\cs}{\mathbb{S}}

\newcommand{\I}{\mathrm{i}}
\newcommand{\dd}{\mathrm{d}}
\newcommand{\mz}{\mathcal{Z}}
\newcommand{\mth}{\mathcal{Z}_{\mathscr{T}}}

\newcommand{\md}{\mathscr{D}}

\newcommand{\mt}{\mathscr{T}}

\newcommand{\tr}{\mathrm{Tr}}
\newcommand{\lieg}{\mathfrak{g}}

\def\e{{\,\rm e}\,}
\newcommand{\ncint}{\int_{\text{supp}(\rho)}\hspace{-1.25cm} - ~ \hspace{0.75cm}}

\usepackage[utf8]{inputenc}
\usepackage[english]{babel}

\usepackage[pdftex]{graphicx}
\usepackage[dvipsnames]{xcolor}

\usepackage[linktocpage=true,colorlinks=true,citecolor=BlueViolet,linkcolor=BlueViolet,urlcolor=BlueViolet]{hyperref}

\usepackage[toc,page]{appendix}
\usepackage[nottoc]{tocbibind}

\usepackage{ytableau}
\usepackage{tikz}
		\usetikzlibrary{er,positioning,arrows.meta}
		\usetikzlibrary{calc}
		\usepackage{tqft}

\begin{document}

{\pagenumbering{roman} 

		\renewcommand*{\thefootnote}{\fnsymbol{footnote}}
	\title{\huge\textbf{$\boldsymbol{T \overline{T}}$-deformation
             of $\boldsymbol q$-Yang-Mills theory}}

	\author[a]{Leonardo Santilli\footnote{lsantilli@fc.ul.pt}}
	\affil[a]{\small Grupo de F\'{\i}sica Matem\'{a}tica, Departamento de Matem\'{a}tica, Faculdade de Ci\^{e}ncias, Universidade de Lisboa, Campo Grande, Edif\'{\i}cio C6, 1749-016 Lisboa, Portugal.}
	
	\author[b,c,d]{Richard J. Szabo\footnote{r.j.szabo@hw.ac.uk}}
	\affil[b]{\small Department of Mathematics, Heriot-Watt University, Colin Maclaurin Building, Riccarton, Edinburgh EH14 4AS, United Kingdom.}
	\affil[c]{\small Maxwell Institute for Mathematical Sciences, Bayes Centre, 47~Potterrow, Edinburgh EH8 9BT, United Kingdom.}
	\affil[d]{\small Higgs Centre for Theoretical Physics, James~Clerk Maxwell Building, Kings Buildings, Edinburgh EH9 3JZ, United Kingdom.}
	
	\author[e,a]{Miguel Tierz\footnote{mtpaz@iscte-iul.pt, tierz@fc.ul.pt}}
	\affil[e]{\small Departamento de Matem\'{a}tica, ISCTE -- Instituto Universit\'{a}rio de Lisboa, Avenida das For\c{c}as Armadas, 1649-026 Lisboa, Portugal.}

	\date{ }
	\maketitle
	\thispagestyle{empty}

        \begin{abstract}
\noindent
We derive the $T\overline{T}$-perturbed version of two-dimensional
$q$-deformed Yang-Mills theory on an arbitrary Riemann surface by
coupling the unperturbed theory in the first order formalism to
Jackiw-Teitelboim gravity. We show that the $T\overline{T}$-deformation results in a
breakdown of the connection with a Chern-Simons theory on a Seifert
manifold, and of the large $N$ factorization into chiral and
anti-chiral sectors. For the $U(N)$ gauge theory on the sphere, we
show that the large $N$ phase transition persists, and that it is of
third order and induced by instantons. The effect of the
$T\overline{T}$-deformation is to decrease the critical value of the
't~Hooft coupling, and also to extend the class of line bundles for
which the phase transition occurs. The same results are shown to hold
for $(q,t)$-deformed Yang-Mills theory. We also explicitly evaluate
the entanglement entropy in the large $N$ limit of Yang-Mills theory,
showing that the $T\overline{T}$-deformation decreases the
contribution of the Boltzmann entropy.
              \end{abstract}

\vspace{2cm}              

\begin{flushright}
  \small{\sf EMPG--20--15}
\end{flushright}

	\clearpage
        {\baselineskip=12pt
          \tableofcontents
          }
}	
	
\bigskip
	
	\pagenumbering{arabic}
	\setcounter{page}{1}
		\renewcommand*{\thefootnote}{\arabic{footnote}}
		\setcounter{footnote}{0}

 \section{Introduction}

Two-dimensional quantum field theories provide a playground for the
study of exactly solvable models, and for testing the relationships and
dualities with other areas such as integrable systems, statistical
mechanics and string theory. Recent broad interest has been attracted
to the solvable irrelevant deformation by the $T\overline{T}$
operator, which is present in any local relativistic two-dimensional quantum field
theory, that was identified by~\cite{Smirnov:2016lqw, Cavaglia:2016oda}
(see~\cite{Jiang:2019hxb} for a review). The novelty of this
deformation is marked by two key properties. First, it does not alter
the integrability of a system. Second, when the field
theory is compactified on a circle, the evolution of the energy levels
with the parameter $\tau$, encoding the strength of the deformation,
is described by a first order inhomogeneous differential equation of
Burgers type. The deformation induced by the operator $T \overline{T}$, henceforth called the
`$T\overline{T}$-deformation', was interpreted in~\cite{Cardy:2018sdv} as a random fluctuation of the background geometry. 
A further step in this direction was made
in~\cite{Dubovsky2017,Dubovsky2018}, where a path integral formulation
of the $T \overline{T}$-deformed theory was put forward: it was proven
in~\cite{Dubovsky2017,Dubovsky2018} that the deformation of a given quantum field theory by the $T\overline{T}$ operator 
is equivalent to coupling the undeformed theory to flat space Jackiw-Teitelboim (JT) gravity. 
These ideas are similar in spirit to the interpretation of the $T\overline{T}$-deformation as a field-dependent spacetime coordinate transformation~\cite{Conti2018geometric}.

After the original formalism for $T\overline{T}$-deformed field theories considered
by~\cite{Smirnov:2016lqw,Cavaglia:2016oda}, a wide range of other
aspects of the $T\overline{T}$-deformation have been
investigated. The original flat space deformation was extended to
two-dimensional quantum field theories on AdS$_2$
in~\cite{Brennan:2020dkw}. Their role in AdS$_3$/CFT$_2$ holography
was investigated
in~\cite{McGough:2016lol,Giveon:2017myj,Hartman:2018tkw,Kraus:2018xrn,Caputa:2019pam,Guica:2019nzm,Lewkowycz:2019xse,Apolo:2019zai,Li:2020pwa}. The
$T\overline{T}$-deformation of Wess-Zumino-Witten (WZW) models was
studied from the string theory perspective in the  
target space theory~\cite{Baggio2018} and also in the gauged worldsheet
sigma-model~\cite{Bonelli2018}. Supersymmetric extensions were
considered
by~\cite{Baggio:2018rpv,Chang:2018dge,Jiang:2019hux,Chang:2019kiu,Ferko:2019oyv,He:2019ahx}. Generalized
$T\overline{T}$-deformations were discussed
in~\cite{LeFloch:2019rut,Conti:2019dxg,Sfondrini20}, while the extensions to
higher-dimensional field theories is considered
by~\cite{Taylor:2018xcy} using holography, and more recently in~\cite{Belin:2020} from direct analysis of the renormalization group flow equation. 
Other facets of the perturbation by the $T \overline{T}$ operator considered recently include the study of the modular properties of the partition
functions of deformed theories~\cite{Datta:2018thy,Aharony:2018bad},
the extension of the $T\overline{T}$-deformation to
non-relativistic systems~\cite{Cardy:2018jho}, and correlation functions in conformal field theories on curved manifolds \cite{Jiang:2019tcq,He:2020udl}. A bridge between the Polyakov loop and the $T \overline{T}$-deformation of a bosonic field theory has been established in \cite{Beratto:2019bap}.\par

In this paper we are concerned with the $T\overline{T}$-deformations
of two-dimensional gauge theories. A simple proposal for the
$T\overline{T}$-deformation of 
Yang-Mills theory on a Riemann surface was advocated
by~\cite{Conti2018}: due to the simple form of the evolution of the
two-dimensional Yang-Mills Hamiltonian with the deformation parameter
$\tau$, the $T\overline{T}$-deformed version of the theory simply
amounts to replacing the quadratic Casimir that appears in the usual heat kernel
expansion of the partition function according to
\begin{align}\label{eq:C2TTbar}
C_2 \longmapsto \frac{C_2}{1-\tau\,\frac{g^2_{\textrm{\tiny YM}}}{2}\,C_2} \ ,
\end{align}
where $g_{\textrm{\tiny YM}}$ is the Yang-Mills coupling
constant. Following this proposal, the phase structure of the $U(N)$
theory on the sphere was studied at large $N$ in~\cite{Santilli2018},
using standard field theory techniques. The 
prescription \eqref{eq:C2TTbar} was derived by~\cite{Ireland} by
directly coupling the heat kernel expansion, which depends on the area
of the Riemann surface, to JT gravity and performing the gravitational
path integral.

The aim of the present paper is to study the analogous features of the
$q$-deformation of two-dimensional Yang-Mills theory. To justify the
effect of the $T\overline{T}$-deformation given by \eqref{eq:C2TTbar},
we follow a different route than~\cite{Ireland}. We consider the first
order formalism for two-dimensional Yang-Mills theory, which rewrites
it as a deformation of BF theory, and hence as an example of an
`almost' topological gauge theory, in the sense which we make precise in
Section~\ref{sec:almosttop}. In this latter general class of theories we can
study the effect of the $T\overline{T}$-deformation precisely through
its coupling to JT gravity, and we reproduce the prescription
of~\cite{Conti2018} as a corollary of a more general result, using
standard abelianisation techniques to evaluate the path integral. At the
same time, thanks to the almost topological nature of these
two-dimensional theories, we can employ cutting and gluing techniques
of topological quantum field theory to rigorously justify the
extension of the $T\overline{T}$-deformation, which is only well-defined on
flat space, to curved Riemann surfaces such as the sphere, which was
not addressed by~\cite{Conti2018,Ireland}.
In a certain sense, the two deformations are compatible: the
$q$-deformation results from a modification of the path integral
measure, leaving the quadratic Casimir unchanged, while the
$T\overline{T}$-deformation modifies only the Hamiltonian, i.e. the
Casimir, but nothing else.

With our techniques, we are able to explore how various facets of
two-dimensional Yang-Mills theory are affected by the
$T\overline{T}$-deformation. For example, we obtain closed expressions
for Wilson loop observables as well as the partition functions on
Riemann surfaces with marked points. However, a number of noteworthy
features are lost under the deformation. For example, the
$T\overline{T}$-deformation of $q$-deformed Yang-Mills theory is no
longer related to Chern-Simons theory (or a deformation thereof) on a
circle bundle over the Riemann surface. Moreover, the large $N$
factorization property, which splits the $U(N)$ gauge theory into chiral and
anti-chiral sectors, no longer holds after deformation. This splitting
is a crucial ingredient in the derivation of the large $N$ string
theory dual of two-dimensional Yang-Mills theory, thus casting doubt
on the existence of such a string theory description of the
$T\overline{T}$-deformed theory. This can be physically understood by
mapping the $T\overline{T}$-deformed gauge theory onto a system of $N$
non-relativistic fermions on a circle, which are now subjected to
non-local interactions leading to long-range correlations between the
fermions.

A central point of this paper is a generalization of the analysis
of~\cite{Santilli2018} to the large $N$ limit of the
$T\overline{T}$-deformation of $q$-deformed $U(N)$ Yang-Mills theory
on the sphere. We show that the main features of the undeformed theory
are preserved, namely there is a third order phase transition induced
by instantons. Furthermore, the $T\overline{T}$-deformation has the
same general features as in the case of ordinary Yang-Mills theory,
and in particular the critical line is lowered as the strength $\tau$
of the deformation is increased. On the other hand, it extends the
class of line bundles for which the phase transition occurs. We also show that these results
continue to hold in the refinement of the theory, known as
$(q,t)$-deformed Yang-Mills theory, whereby the region of the small
coupling phase is reduced by the refinement.

The exact solvability of two-dimensional Yang-Mills theory also makes
it an interesting testing ground for exploring concepts of quantum
information, such as quantum entanglement, in quantum field
theory. Generally, the entanglement entropy of codimension one spatial
subregions provides a powerful tool for investigating new perspectives
in quantum field theory, but it is notoriously difficult to
compute. Most examples involve only free fields or explicit conformal
symmetry, whereas the highly interesting examples involve the
renormalization group (RG) flow of the entanglement entropy. For the
RG flow triggered by the $T\overline{T}$-deformation,
the entanglement entropy was computed in~\cite{Ireland} for finite
$N$, where it was found that the effect of the deformation is
relatively mild, which is anticipated from the ultraviolet finiteness
of two-dimensional Yang-Mills theory. In this paper we study the
entanglement entropy of the
$T\overline{T}$-deformed gauge theory at large $N$, where we find that the
contribution from the Shannon entropy vanishes, while the contribution
from the Boltzmann entropy, per point of the entangling surface, is
explicitly evaluated and shown to decrease as the strength $\tau$ of
the deformation is increased.

\paragraph{Organization of the paper.}
	In Section \ref{sec:almosttop} we present our formalism for
        generic almost topological gauge theories. In Section
        \ref{sec:TTqYMfinite} we then focus on Yang-Mills
        theory in two dimensions together with its $q$-deformation and
        subsequent refinement which depend, among other continuous
        moduli, on a discrete parameter $p\in\Z$; we present their $T
        \overline{T}$-deformation and study how their well-known
        properties are changed by the deformation. Section
        \ref{sec:phaseTTqYM} is dedicated to the study of the phase
        structure at large $N$, where we find that the expected phase
        transition extends to $p<2$ as a consequence of the
        $T\overline{T}$-deformation. In Section \ref{sec:entanglement}
        we present some results for the entanglement entropy of these
        theories. We conclude with possible avenues for future
        research in Section \ref{sec:out}. Two appendices at the end
        of the paper contain some technical details that supplement
        the analyses of the main text.
	
\paragraph{Conventions.}
	To avoid excessive repetition of the word `deformation', we
        will only explicitly state it when using the terminology `$T \overline{T}$-deformation'. The
        $q$-deformed Yang-Mills theory and its refinement,
        $(q,t)$-deformed Yang-Mills theory, will be henceforth simply called
        `$q$-Yang-Mills theory' and `$(q,t)$-Yang-Mills theory', respectively.
	
	\section{$\boldsymbol{T \overline{T}}$-deformation of almost
          topological gauge theories}
	\label{sec:almosttop}
        
		Consider the partition function of a gauge theory
                $\mt$ with compact connected gauge group $G$ on a Riemann surface
                $\Sigma$ which is described by the insertion of a non-local
                operator $\mathcal{O}(\Phi)$ in the path integral of a two-dimensional topological quantum field theory:
		\begin{equation*}
			\mth [ \Sigma ] = \int\, \md \Phi~ \e^{-
                          S_{\textrm{\tiny TQFT} } ( \Phi ) } \,
                        \mathcal{O} (\Phi) =: \big\langle \mathcal{O}
                        (\Phi)  \big\rangle_{\textrm{\tiny TQFT}} \ .
		\end{equation*}
		Here $\Phi$ collectively denotes the fields of the
                theory and $\md \Phi$ is a gauge-invariant measure on
                the space of fields, while the action
                $S_{\textrm{\tiny TQFT} }(\Phi) $ defines a topological
                field theory. Notable examples of such theories are
                two-dimensional Yang-Mills theory and its relatives,
                which arise from a BF-type topological gauge theory 
                through a deformation that is precisely of this type,
                as we will review in
                Section~\ref{sec:TTqYMfinite}. These theories will be
                the focus of subsequent sections. Nevertheless, one
                may also consider deformations of the topologically
                twisted sigma-models of
                \cite{Witten:1988xj,Witten1991mirror} or of other
                classes of two-dimensional topological field theories
                \cite{Malik:2000yi} by some non-local operator, and
                our considerations in this section also pertain to
                these more general gauge theories.
                
		The spacetime, on which our field theory is defined,
                is a Riemann surface $\Sigma$, possibly with
                $s$ marked points decorated with representations $R_1
                , \dots , R_s$ of the gauge group $G$, in which case
                the partition function is denoted by 
		\begin{equation*}
			\mth [ \Sigma ; R_1 , \dots , R_s ] \ .
		\end{equation*}
		The surface $\Sigma$ is allowed to have boundaries,
                and the partition function will be understood as a
                function of suitable boundary conditions, which in
                particular include the holonomies of the gauge
                connection along the one-dimensional boundaries. Field
                theories without gauge symmetries can be considered as
                well in this framework as a special case with trivial
                gauge group.
                
		Theories of this class are amenable to the $T
                \overline{T}$-deformation, albeit defined on a curved
                spacetime $\Sigma$, thanks to their ``almost''
                topological nature. According to
                \cite{Dubovsky2017,Dubovsky2018} (see also
                \cite{Freedman2019}), $T \overline{T}$-deformation is
                equivalent to coupling the field theory to
                two-dimensional topological gravity. If the theory we
                start with is topological, the gravitational sector of
                the path integral can be integrated out with no
                effect. However, if non-local operators have been
                inserted, they couple to the gravitational sector and
                the $T\overline{T}$-deformation is represented
                symbolically as
		\begin{equation*}
			\mth [ \Sigma ] \ \xrightarrow{ \ T \overline{T}\text{-deformation} \ }  \   \mth ^{T \overline{T}} [ \Sigma ] 
		\end{equation*}
		with 
		\begin{equation*}
			 \mth ^{T \overline{T}} [ \Sigma ]
                         = \left\langle \frac{1}{\mz _{\textrm{\tiny
                                 JT} } }\, \int\, \md \boldsymbol{e} ~
                           \mathcal{O} (\Phi ; \boldsymbol{e} ) ~
                           \exp\Big(\mbox{$\frac{1}{2 \tau}\,
                             \int_{\Sigma}\,
                             (\boldsymbol{e}-\boldsymbol{f}) \wedge
                             (\boldsymbol{e}-\boldsymbol{f})$}\Big)
                         \right\rangle_{\textrm{\tiny TQFT}} \ .
		\end{equation*}
This is the path integral of JT gravity, normalized by the pure
gravity partition function $\mathcal{Z}_{\textrm{\tiny JT}}$. The
integration is over the coframe field $\boldsymbol{e}$ of the target space,
with $\boldsymbol{f}$ the coframe field of the worldsheet $\Sigma$,
and the path integral measure is induced by the metric
\begin{align*}
\delta s^2 = \int_\Sigma\,
  \delta\boldsymbol{e}\wedge\delta\boldsymbol{e} \ .
\end{align*}
The
notation $\mathcal{O} (\Phi ; \boldsymbol{e} )$ means that the
non-local operator has the same form as before, but now lives in the
manifold with coframe field $\boldsymbol{e}$. For derivative-free operators
$ \mathcal{O} (\Phi)$ this simply means that, in every integral, we
have to replace the original volume form $\omega$ on $\Sigma$, written
in terms of the
coframe field of $\Sigma$ as $\boldsymbol{f} \wedge \boldsymbol{f}$, by the target space volume form $\boldsymbol{e}
\wedge \boldsymbol{e}$. This presentation is equivalent to the change
of variables described in \cite{Conti2018geometric}, but for the
purposes of the present work we find it convenient to use the explicit
path integral presentation.

		The proof of equivalence with the $T
                \overline{T}$-deformation presented
                in~\cite{Dubovsky2018} relies on showing that the gravitational path
                integral is one-loop exact, and reproduces the
                $T\overline{T}$-deformed partition function. This ties
                in nicely with the arguments of~\cite{Dubovsky2017}
                that the gravitational dressing provided by the
                $T\overline{T}$-deformation is a semi-classical effect. We shall see
                this explicitly for the class of non-local operators
                that we ultimately consider below.
		
		We always consider the surface $\Sigma$ to be equipped
                with a Riemannian metric, and therefore use the
                Euclidean gravity path integral, following the
                conventions of \cite{Freedman2019} (which agree with
                those of \cite{Conti2018,Conti2018geometric}). The
                deformation parameter $\tau$ in the present paper then
                differs by a sign from the conventions of
                \cite{Dubovsky2018,Tolley2019} which work in
                Lorentzian signature. 
		
		It was noted in \cite{Ireland} (see also
                \cite[Appendix~A]{Tolley2019} for relevant discussion)
                that there is a subtlety in the normalization of the
                path integral measure $\mathscr{D} \boldsymbol{e}$: as
                will be manifest below, assuming the naive
                normalization of the measure and performing the
                gravity path integral, one does not recover the
                undeformed theory in the limit $\tau \to 0$. Imposing
                the latter condition instead leads to a choice of
                normalization for the path integral measure $\md
                \boldsymbol{e}$ which depends on $\mathcal{O}
                (\Phi)$. In particular, the order of the path
                integrations do not commute, and the topological
                gravity degrees of freedom should always be introduced
                inside the correlator $\langle \,\cdot\,
                \rangle_{\textrm{\tiny TQFT}}$. Below we will provide
                a more extensive comparison between the present
                analysis and that of \cite{Ireland}, and this
                technical aspect will play an important role. 
		
		With the application to two-dimensional Yang-Mills
                theory along with its generalizations and deformations
                in mind, we now specialize our analysis to the case in
                which the functional dependence of $\mathcal{O}
                (\Phi)$ is through operators of the form
		\begin{equation*}
			\mathcal{O} (\Phi) = \exp \left( -
                          \frac{\lambda}{2 N} \, \int_{\Sigma}\,
                          V(\phi, \psi ) ~ \omega \right) \ ,
		\end{equation*}
		where $\omega$ is the normalized volume form on $\Sigma$
                and the potential $V (\phi, \psi)$ is a scalar functional
                of scalar fields $\phi$ and possibly spinor fields
                $\psi$. The coupling is $\frac{\lambda}{N}$, where
                $\lambda$ is a 't Hooft parameter and $N$ is the rank
                of the gauge group $G$. When coupled to
                topological gravity, this operator becomes
		\begin{equation*}
			\mathcal{O} (\Phi ; \boldsymbol{e}) = \exp
                        \left( - \frac{\lambda}{2 N}\, \int_{\Sigma}\,
                          V(\phi, \psi ) ~ \boldsymbol{e} \wedge
                          \boldsymbol{e} \right) \ ,
		\end{equation*}
		and the $T \overline{T}$-deformed partition function reads 
		\begin{equation*}
			\mth ^{T \overline{T}} [ \Sigma ]  = \int\,
                        \md \Phi \ \frac{1}{\mz_{\textrm{\tiny JT}}}
                        \, \int\, \md \boldsymbol{e}\, \exp\Big(\mbox{$-
                          S_{\textrm{\tiny TQFT} } ( \Phi )  -
                          \lambda\, \int_{\Sigma}\, \big( \frac{1}{2N}\,
                          V(\phi, \psi )\ \boldsymbol{e} \wedge
                          \boldsymbol{e} - \frac{1}{2 \tau}\, (
                          \boldsymbol{e} - \boldsymbol{f} ) \wedge  (
                          \boldsymbol{e} - \boldsymbol{f} ) \big)  $} \Big) \, .
		\end{equation*}
		We have chosen a non-standard definition of the parameter $\tau$, including an overall factor $\lambda$, which will be convenient in the forthcoming discussion. After simple manipulation and integrating over $\boldsymbol{e}^{\prime} = \boldsymbol{e} - \boldsymbol{f}$, one obtains 
		\begin{align}
			\mth ^{T \overline{T}} [ \Sigma ]  & =  \int\,
                                                             \md \Phi
                                                             \
                                                             \exp\Big(\mbox{$-
                                                             S_{\textrm{\tiny
                                                             TQFT} } (
                                                             \Phi )  -
                                                             \frac{\lambda}{2}\,
                                                             \int_{\Sigma}\,
                                                             N^{-1}\,
                                                             V (\phi,
                                                             \psi )\
                                                             \boldsymbol{f}
                                                             \wedge
                                                             \boldsymbol{f}$} \notag
                                                             \\ &
                                                                  \hspace{3cm}
                                                                  \left. \mbox{$+ \frac\lambda2\,\int_\Sigma\,
                                                                  \frac{
                                                                  \tau\,
                                                                  N^{-1}
                                                                  }{ 1
                                                                  -
                                                                  \tau\,
                                                                  N^{-1}\,
                                                                  V
                                                                  (\phi,
                                                                  \psi)
                                                                  } \big(  N^{-1}\, V (\phi, \psi )\, \boldsymbol{f} \big) \wedge \big(  N^{-1}\, V (\phi, \psi)\, \boldsymbol{f} \big) $}\right) \notag \\[4pt]
				& =  \int\, \md \Phi \
                           \exp\left(\mbox{$- S_{\textrm{\tiny TQFT} }
                           ( \Phi ) - \frac{\lambda}{2 N}\,
                           \int_{\Sigma}\, \frac{ V (\phi, \psi )}{ 1
                           - \frac{\tau}{N}\, V (\phi , \psi) }\,
                           \omega $}\right) \ ,  \label{eq:TTdefVPI}
		\end{align}
		which correctly reproduces the prescription of
                \cite{Conti2018} for the $T\overline{T}$-deformation
                given by
		\begin{equation}
		\label{eq:Vttdef}
			\frac{\lambda}{2 N}\, V (\phi, \psi )
                        \longmapsto \frac{ \frac{\lambda}{2 N }\, V (
                          \phi, \psi )}{ 1 - \frac{\tau}{ N}\, V
                          (\phi, \psi )} \ .
		\end{equation}
		In \eqref{eq:TTdefVPI} we use a $V (\phi,
                \psi)$-dependent normalization of the gravity path
                integral that cancels a factor from the Gaussian
                integration. Had we not done so, we would not recover
                the undeformed partition function in the limit $\tau \to 0$. This is the incarnation of the subtleties in the normalization of the path integral measure discussed in \cite{Ireland,Tolley2019}.
				
		\subsubsection*{$\boldsymbol{T \overline{T}}$-deformation in curved
                  spacetime}
                
		The $T \overline{T}$-deformation is only well-defined
                in flat space. However, in the present setting, we 
                argue that, since the underlying theory is
                topological, for suitable insertions $\mathcal{O}
                (\Phi)$ we can define the $T \overline{T}$-deformation
                on flat space, and then put the theory on a curved
                manifold $\Sigma$. We now explain this point more
                rigorously.
                
\begin{figure}[htb]
		\centering
			\begin{tikzpicture}[tqft/flow=east,node distance=4cm]
			\begin{scope}[tqft/boundary style=draw]
			\node[tqft/pair of pants,draw] (a) at (4,0) {};
			\node[tqft/cylinder,draw] (b) at (0,0) {};
			\node[tqft/cap,draw] (c) at (-4,0) {};
			\end{scope}
			\end{tikzpicture}
		\caption{\small The disk (left), the cylinder (center) and the pair of pants (right), homeomorphic to the complex plane with respectively zero, one or two holes.}
		\label{fig:buildingTQFT}
		\end{figure}
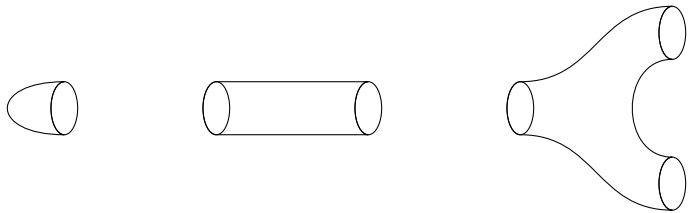

		Thanks to the cutting and gluing property of
                topological quantum field theories
                \cite{atiyah1988topological,segal1988definition}, one
                can decompose $\Sigma$ into disks, cylinders and pairs
                of pants, obtaining the same theory on each piece (see
                Figure \ref{fig:buildingTQFT}). Such components have
                boundaries, and one should impose suitable boundary
                conditions on the fields. The disk, the cylinder and
                the pair of pants are homeomorphic respectively to the
                complex plane $\mathbb{C}$, the punctured plane
                $\mathbb{C}^{\times}$ and the doubly-punctured plane $\C^{\times\times}$. 
		Therefore, we reduce the topological quantum field
                theory on flat components, which are many copies of
                the complex plane $\mathbb{C}$, with either zero, one
                or two holes. At this point, we insert $\mathcal{O}
                (\Phi)$ on each component, and turn on gravity. Since
                each component is flat, the $T
                \overline{T}$-deformation prescription is well-defined
                on each component.

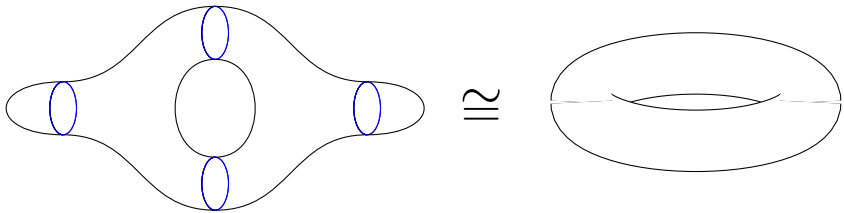
\begin{figure}[htb]
		\centering
			\begin{tikzpicture}[tqft/flow=east]
			\begin{scope}[tqft/boundary upper style={draw,blue},tqft/boundary lower style={draw,blue}]
			\node[tqft/pair of pants,draw] (a) at (0,0) {};
			\node[tqft/cap,draw,anchor=outgoing boundary 1] (b) at (a.incoming boundary 1) {};
			\node[tqft/reverse pair of pants,draw] (c) at (2,0) {};
			\node[tqft/cup,draw,anchor=incoming boundary 1] (d) at (c.outgoing boundary 1) {};
			\end{scope}
			\node[] (eq) at (4.5,0) {\huge $\cong$};
			\end{tikzpicture}%
			\begin{tikzpicture}[yscale=cos(70)]
			\draw[double distance=8mm] (0:1.5) + (0.5,3) arc (0:180:1.5) + (0.5,3);
   			\draw[double distance=8mm] (180:1.5) + (0.5,3) arc (180:360:1.5) + (0.5,3);
			\end{tikzpicture}
		\caption{\small Obtaining a torus from elementary pieces. On the left, the gluing is an integration over boundary conditions (in blue).}
		\label{fig:torusglue}
		\end{figure}

		After performing the JT gravity path integral, we can
                glue back together the pieces and reassemble $\Sigma$
                (see Figure \ref{fig:torusglue}). Topological gravity
                couples to bulk geometry and does not change the
                boundary data, at least for operator insertions
                $\mathcal{O} (\Phi)$ of the form in \eqref{eq:TTdefVPI} (we will briefly comment on the most general case shortly). Hence the gluing goes exactly as without $T \overline{T}$-deformation, and we obtain a $T \overline{T}$-deformed theory on $\Sigma$.
		
		At this point it is worthwhile mentioning the proposal \cite{Tolley2019} that $T \overline{T}$-deformation in curved spacetime corresponds to massive gravity. In the present setting, we could equivalently take the (Euclidean version of the) proposal of \cite{Tolley2019} as the definition of $T \overline{T}$-deformation on curved two-dimensional manifolds.
		A step towards a rigorous definition of generic $T
                \overline{T}$-deformed theories on curved manifolds
                has also been taken in \cite{Shyam:curved}.

		\subsubsection*{Boundary quantum mechanics}
                
			In the procedure of putting the theory on
                        $\Sigma$, we exploited the underlying
                        topological quantum field theory and a
                        suitable form of the insertion $\mathcal{O}
                        (\Phi)$. However, a generic operator
                        $\mathcal{O} (\Phi )$ may in principle couple
                        dynamically to the boundary conditions,
                        spoiling the gluing technique. We find it
                        appropriate to say a few more words about this
                        more general scenario, although it will play
                        no role in the rest of our discussion.
                        
			First, we notice that there will never be any one-dimensional interfaces, since we insert the same operator on each two-dimensional component. 
			In the general setting, one may need to deal
                        with a dynamical one-dimensional boundary
                        theory. Thinking of the $T
                        \overline{T}$-deformation as a field-dependent
                        change of variables \cite{Conti2018geometric}, we expect the $T
                        \overline{T}$-deformed boundary theory to be a
                        rewriting of the original theory in the new
                        variables, at least at the classical level. We
                        notice that no interface will arise in any
                        case, since we are putting the same operator $\mathcal{O} (\Phi)$ in each component. 
			Therefore we should be able, at least in
                        principle, to glue the pieces together, at the
                        price of solving the gluing theory~\cite{Dedushenko:2018aox}, which will be a
                        quantum mechanics deformed by the effect of
                        the change of variables on the boundary modes.
                        
	\section{Two-dimensional Yang-Mills and $\boldsymbol q$-Yang-Mills theories}
	\label{sec:TTqYMfinite}
        
In this paper we study two-dimensional Yang-Mills theory, its
$q$-deformation and its subsequent refinement to two-dimensional
$(q,t)$-deformed Yang-Mills theory. These are examples of almost
topological gauge theories of the type discussed in
Section~\ref{sec:almosttop}, where the underlying topological field
theory is two-dimensional BF theory,\footnote{This theory is sometimes
also refered to as two-dimensional topological Yang-Mills theory.} whose fields are a scalar field
$\phi$ on $\Sigma$ in the adjoint representation of the gauge algebra
$\lieg$ and the curvature $F^A$ of a gauge connection $A$ on (a
trivial principal $G$-bundle over)
$\Sigma$. Ordinary Yang-Mills theory on $\Sigma$ corresponds to a
deformation of this BF theory by a non-local operator $\mathcal{O}(\phi)$ which
adds a potential $V(\phi)=\tr\,\phi^2$ to the BF action. This theory
can be $\beta$-deformed by modifying the discrete matrix model which
arises for $\beta=2$ to a general $\beta$-ensemble. One can further
deform the underlying BF theory by making the field $\phi$ compact, that is, taking
it to be valued in the adjoint representation of the gauge group
$G$. Adding the potential $V(\phi)$ deforms this theory to
$q$-Yang-Mills theory which can be subsequently refined to
$(q,t)$-Yang-Mills theory, that is a categorification of the
$\beta$-ensemble. The initial theories, with non-compact $\phi$, can
then be regarded as classical limits $q\to1$ of the theories with
compact scalar~$\phi$.

We depict these relationships between the various incarnations of
Yang-Mills theory on $\Sigma$ through the diagram
	\begin{equation*}
		\begin{tikzpicture}[auto,node distance=3cm]
			\node[] (BF2) {BF$_2$};
			\node[] (BFq) [below = of BF2] {compact BF$_2$};
			\node[] (YMq) [right = of BFq] {$q$-YM$_2$};
			\node[] (YM2) [above = of YMq] {YM$_2$};
			\node[] (tqYM) [right = of YMq] {$(q,t)$-YM$_2$};
			\node[] (bYM) [above = of tqYM] {$\beta$-YM$_2$};
			\path[->] (BF2) edge node {\footnotesize add potential $\phi^2$} (YM2);
			\path[->] (BFq) edge node {\footnotesize add potential $\phi^2$} (YMq);
			\path[->] (BF2) edge [left] node {\footnotesize compact $\phi$} (BFq);
			\path[->] (YM2) edge [bend right,left] node {\footnotesize compact $\phi$} (YMq);
			\path[->] (YMq) edge [bend right,right] node {\footnotesize $q \to 1$} (YM2);
			\path[->] (YM2) edge node {\footnotesize $\beta$-deformation} (bYM);
			\path[->] (YMq) edge node {\footnotesize refinement} (tqYM);
			\path[->] (tqYM) edge [bend right,right] node {\footnotesize $q \to 1$} (bYM);
			\path[->] (bYM) edge [bend right,left] node {\footnotesize categorify} (tqYM);
		\end{tikzpicture}
	\end{equation*}
In this section we use the formalism developed in Section
\ref{sec:almosttop} to study the $T \overline{T}$-deformation of the
Yang-Mills theories appearing in this diagram. 
	
	\subsection{$T \overline{T}$-deformation of two-dimensional
          Yang-Mills theory}
        \label{sec:TTbarYM}
        
	In \cite{Conti2018} the $T \overline{T}$-deformation of
        two-dimensional Yang-Mills theory on $\Sigma$ was obtained, through explicit solution of the flow equation
	\begin{equation*}
		\frac{ \partial \mathcal{L}(\tau)}{
                  \partial \tau} = \det_{\mu, \nu =1,2}\, \big[T_{\mu \nu}(\tau)\big]
                \  ,
	\end{equation*}
	to all orders in $\tau\in[0,\infty)$. This equation is to be
        solved with the initial condition on
        the deformed Lagrangian $\mathcal{L}(\tau)$ that requires $\mathcal{L}(0)=\mathcal{L}_{\textrm{\tiny
            YM}}$ to be  the Yang-Mills Lagrangian, and
        $T_{\mu\nu}(\tau)$ is the Hilbert energy-momentum tensor of the two-dimensional field theory. The same
        deformation has recently been obtained in \cite{Ireland}
        through the coupling with JT gravity. We rederive the result
        by exploiting the equivalent first order formulation as a
        deformation of BF theory. The argument is as follows:
        Yang-Mills theory is a pure gauge theory, but it is equivalent to a
        BF theory with additional Gaussian term for the scalar $\phi\in\Omega^0(\Sigma,\lieg)$
        given by (see \cite{BT93lectures})
	\begin{equation*}
		S_{\textrm{\tiny YM}} = \frac{N}{2 \lambda}\,
                \int_{\Sigma}\,\tr\, F^{A} \ast F^{A} = \int_{\Sigma}\,\tr
                \Big(  \mathrm{i}\, \phi\, F^{A} + \frac{\lambda}{2
                  N}\, \phi^2\, \omega\Big) \ ,
              \end{equation*}
              where the equality is understood to hold on-shell.
Here $F^{A}\in\Omega^2(\Sigma,\lieg)$ is the curvature of a gauge connection $A$ on $\Sigma$, $\tr$ is an
invariant quadratic form on the Lie algebra $\lieg$, $\omega$ is the
symplectic structure on $\Sigma$ and $\ast$ is the Hodge operator
constructed from the Riemannian metric compatible with $\omega$. The Yang-Mills coupling is
	\begin{equation*}
		g_{\textrm{\tiny YM}} ^2 = \lambda\, N^{-1} \ .
              \end{equation*}
        The first term is the action of two-dimensional BF theory
        which is topological, thus the $T
        \overline{T}$-deformation of the first order formulation only
        changes the potential $V(\phi) = \tr\,
        \phi^2$ as in \eqref{eq:Vttdef}.
	From this point, the derivation of the heat kernel expansion using
        abelianization of the path integrals goes exactly as
        in~\cite{BT93lectures}: one conjugates the scalar field $\phi$
        into a Cartan subalgebra of $\lieg$ using gauge invariance
        and the Weyl integral formula, and then integrates over the root
        components $A_\alpha$ of the gauge connections with respect to the root
        space decomposition of the Lie algebra $\lieg$. Then two-dimensional Yang-Mills theory can
        be $T \overline{T}$-deformed by replacing the quadratic Casimirs
        of representations $R$ of $G$ according to\footnote{We are slightly changing the normalization of the topological gravity action, $\frac{1}{\tau} \mapsto \frac{N^2}{\tau}$, to make the right-hand side well-defined at all $\tau$ for every $N$.}
	\begin{equation}
	\label{eq:C2toTT}
		C_2 (R) \longmapsto C_2 ^{T \overline{T}} (R,\tau) :=
                \frac{C_2 (R) }{ 1 - \frac{\tau}{N^3}\, C_2 (R) }  \ .
              \end{equation}
              
	This derivation immediately extends to the generalized
        Yang-Mills theory of \cite{Ganor1994}, where higher order
        Casimir operators are included by adding higher degree terms
        to the potential $V(\phi)$. These can include multi-trace
        terms, since the derivation does not rely on the explicit form
        of $V(\phi)$. The $T \overline{T}$-deformation of the
        generalized two-dimensional Yang-Mills theory is then directly
        obtained from \eqref{eq:Vttdef}. The final answer for the
        partition function of generalized Yang-Mills theory is then
	\begin{equation}
	\label{eq:genTTYM2}
		\mz_{\textrm{\tiny gen-YM}} ^{T \overline{T}} [\Sigma]
                = \sum_{R}\, \dim (R)^{\chi (\Sigma)} \, \exp\Big(- \frac{\lambda}{2N}\, \frac{ C_{\text{gen}} (R) }{1 -  \frac{\tau}{N^3}\, C_{\text{gen}} (R) } \Big) \ ,
	\end{equation}
	where $C_{\text{gen}}(R) $ includes the quadratic and higher
        Casimir operators. The sum runs over isomorphism classes of
        irreducible representations $R$ of $G$ \cite{CMRlectures} with
        dimension $\dim(R)$, the coupling $\lambda$ is identified with
        the area $A$ of the surface $\Sigma$, and $\chi(\Sigma)$ is the
        Euler characteristic of $\Sigma$.
	
	\subsubsection*{Comparison with the literature}
        
	Since the partition function of $T \overline{T}$-deformed
        two-dimensional Yang-Mills theory has been derived in
        different ways in the literature
        \cite{Conti2018,Ireland,Brennan:2019azg}, it is appropriate to now
        pause and discuss our result.
        
	Formula \eqref{eq:genTTYM2}, or more precisely its original
        version with $C_{\text{gen}}(R) = C_2 (R)$, was first proposed in~\cite{Conti2018}, although it was not rigorously justified for
        curved surfaces $\Sigma$. The proposal of \cite{Conti2018} was
        also the starting point of previous work \cite{Santilli2018}
        studying the phase structure of the $T\overline{T}$-deformed
        theory. Here we have provided the derivation, following an
        argument similar to that of \cite{Ireland} but with a few
        important differences.
        
	In \cite{Ireland} topological gravity is introduced after integrating out the gauge fields. In particular, as carefully explained there, the JT gravity path integral is representation-dependent and is inserted inside the sum over irreducible representations. Schematically 
	\begin{equation*}
		\sum_{R}\, Z_R (\omega ) \xrightarrow{ \ \text{$T
                    \overline{T}$-deformation of \cite{Ireland}} \ }
                \sum_{R}\, \int\, \mathscr{D}_R \boldsymbol{e} ~ Z_R
                (\boldsymbol{e} \wedge \boldsymbol{e} ) \ ,
	\end{equation*}
	where $Z_R (\omega)$ is the summand in \eqref{eq:genTTYM2},
        and we have stressed its dependence on the volume form
        $\omega$ of $\Sigma$. The normalization of the measure
        $\mathscr{D}_R \boldsymbol{e}$ is taken to be $R$-dependent.
        
	Therefore, the procedure of \cite{Ireland} does not deform the
        original path integral, but deforms each summand in the
        expression obtained after abelianization
        \cite{BT93lectures}. The technique we adopted, instead,
        describes a deformation of the full path integral, and proves
        that the abelianization takes place also in the $T
        \overline{T}$-deformed theory. The two results coincide, as
        expected. Indeed the gauge fields do not enter in the
        definition of the operator $\mathcal{O} (\phi)$, which couples
        to gravity. For this reason, the integration over the coframe field
        is expected to commute with the integration over the gauge
        fields. 
	
	\subsection{$T \overline{T}$-deformed $q$-Yang-Mills theory}
        \label{sec:TTbarqYM}
        
	We can extend the argument above to $q$-deformed Yang-Mills
        theory: this deformation modifies the domain of integration,
        making the scalar field $\phi$ compact, i.e. taking
        $\phi\in\Omega^0(\Sigma,G)$ to be valued in the Lie
        group $G$ instead of its Lie algebra $\lieg$, without altering the
        action~\cite{Aganagic2004,BT06}. In this case abelianization
        proceeds by conjugating $\phi$ into the maximal torus of $G$. The $T
        \overline{T}$-deformation thus changes the potential for the
        (now compact) scalar, exactly as in the case of ordinary
        two-dimensional Yang-Mills theory. The final answer for the $T
        \overline{T}$-deformed partition function of $q$-Yang-Mills
        theory on a surface $\Sigma$ of genus $g_{\Sigma}$ with $s$
        boundaries is 
	\begin{equation}
	\label{eq:ZqYMTT}
		\mz_{\textrm{\tiny $q$-YM}} ^{T \overline{T}} [\Sigma;
                g_1, \dots, g_s] = \sum_{R}\, \dim_q (R)^{\chi
                  (\Sigma)} \, q^{\frac{p}{2}\, C_2 ^{T \overline{T}}
                  (R,\tau)  } \ \chi_{R} (g_1) \cdots   \chi_{R} (g_s) \ ,
	\end{equation}
	with the identification of the $q$-parameter
        $$
        q= \e^{-\lambda
          /N } \ .
        $$
        Here $p\in\Z$ is a discrete parameter, the $T \overline{T}$-deformed Casimir is defined in \eqref{eq:C2toTT}, and 
	\begin{equation*}
		 \chi (\Sigma) = 2 - 2 g_{\Sigma} - s
	\end{equation*}
	is the Euler characteristic of $\Sigma$. The boundary
        conditions $g_1, \dots, g_s\in G$ are the holonomies of the
        gauge connection around the boundaries, with characters
        $\chi_R$ in the representation $R$, and $\dim_q(R)$ is the
        quantum dimension of $R$. For closed surfaces $\Sigma$, the
        formula \eqref{eq:ZqYMTT} is simply 
	\begin{equation*}
		\mz_{\textrm{\tiny $q$-YM}} ^{T \overline{T}} [\Sigma]
                = \sum_{R}\, \dim_q (R)^{2 - 2 g_{\Sigma}} \,
                q^{\frac{p}{2}\, C_2 ^{T \overline{T}} (R,\tau)  } \ .
	\end{equation*}
	Again, the argument straightforwardly extends to generalized
        $q$-deformed Yang-Mills theory, with additional higher degree
        terms added to the potential $V(\phi)$.
        
	The partition function of the ordinary $T
        \overline{T}$-deformed Yang-Mills theory from
        Section~\ref{sec:TTbarYM} above is recovered by taking the limit 
	\begin{equation}
	\label{eq:qtoYMlimit}
		p \to \infty \qquad \mbox{and} \qquad \lambda \to 0
                \qquad \mbox{with} \quad \lambda\, p = A ~
                \text{fixed}  \ ,
	\end{equation}
	where $A$ is the area of $\Sigma$.
	
	\subsubsection*{$\boldsymbol q$-Yang-Mills theory on the disk and on the
          cylinder}

		As we have shown, the procedure of
                $T\overline{T}$-deformation works for every Riemann
                surface $\Sigma$, possibly with
                boundary.\footnote{Two-dimensional Yang-Mills theory
                  can also be defined on surfaces with corners
                  \cite{YM2Corners,Iraso2019}, and it is likely that our technique extends to that case.} Special roles are played by the disk and cylinder partition functions. On the disk we have 
		\begin{equation*}
	    	\mz_{\textrm{\tiny $q$-YM}} ^{T \overline{T}} \left[ \begin{tikzpicture}[tqft/flow=east,baseline=-0.5ex]
			\begin{scope}[tqft/boundary style=draw]
			\node[tqft/cap,draw] (c) {};
			\end{scope}
			\end{tikzpicture} ~ 
			; g \right] = \sum_{R}\, \dim_q (R)  ~q^{\frac{p}{2}\,  C_2 ^{T \overline{T}} (R,\tau) } ~\chi_{R} (g) \  .
		\end{equation*}
		Gluing two disks whose boundaries have opposite
                orientations and using the orthogonality of the
                characters we get the $T \overline{T}$-deformed
                partition function on the sphere $\mathbb{S}^2$: 
		\begin{align*}
			\int_G\, \dd g ~ \mz_{\textrm{\tiny $q$-YM}} ^{T \overline{T}} \left[ \begin{tikzpicture}[tqft/flow=east,baseline=-0.5ex]
			\begin{scope}[tqft/boundary style=draw]
			\node[tqft/cap,draw] (c) {};
			\end{scope}
			\end{tikzpicture} ~ 
			; g \right]  ~ \mz_{\textrm{\tiny $q$-YM}} ^{T \overline{T}} \left[ \begin{tikzpicture}[tqft/flow=east,baseline=-0.5ex]
			\begin{scope}[tqft/boundary style=draw]
			\node[tqft/cup,draw] (c) {};
			\end{scope}
			\end{tikzpicture} ~ 
			; g^{-1} \right] = \mz_{\textrm{\tiny $q$-YM}} ^{T \overline{T}} \left[ \begin{tikzpicture}[tqft/flow=east,baseline=-0.5ex]
			\begin{scope}[tqft/boundary upper style={draw,blue},tqft/boundary lower style={draw,blue}]
			\node[tqft/cap,draw] (c) {};
			\node[tqft/cup,draw,anchor=incoming boundary 1] (d) at (c.outgoing boundary 1) {};
			\end{scope}
			\end{tikzpicture} \right] = \mz_{\textrm{\tiny
                  $q$-YM}} ^{T \overline{T}} \left[ \mathbb{S}^2
                  \right]  \ ,
		\end{align*}
                where $\dd g$ is the invariant Haar measure on $G$.
		
		The cylinder partition function is 
		\begin{equation*}
	    	\mz_{\textrm{\tiny $q$-YM}} ^{T \overline{T}} \left[ \begin{tikzpicture}[tqft/flow=east,baseline=-0.5ex]
			\begin{scope}[tqft/boundary style=draw]
			\node[tqft/cylinder,draw] (c) {};
			\end{scope}
			\end{tikzpicture} ~ 
	    	; g_{\text{in}}, g_{\text{out}} \right] = \sum_{R}\, q^{\frac{p}{2}\,  C_2 ^{T \overline{T}} (R,\tau) } ~\chi_{R} (g_{\text{in}}^{-1})\, \chi_{R} (g_{\text{out}}) \  ,
		\end{equation*}
		where we have already taken into account the
                orientation in the definition of the boundary
                condition $g_{\text{in}}$. In the topological limit $\lambda=0$, it serves as a propagator: attaching it to any surface $\Sigma$ replaces the holonomy $g_{\text{in}}$ by $g_{\text{out}}$. At non-zero area though, attaching a cylinder has a notable effect which effectively increases the coupling. With our choice of normalization for $\tau$, the effect of gluing a cylinder to $\Sigma$ is precisely the same as in the theory without $T \overline{T}$-deformation.
		
                \subsubsection*{Supersymmetry}
                
		An additional consistency check for our formulas comes
                from the minimal supersymmetric extension of
                Yang-Mills theory. Two-dimensional Yang-Mills theory
                and its $q$-deformation are equivalent to their
                supersymmetric counterparts. The BRST multiplet is
                $(A, \phi, \psi)$, with $\psi$ a Grassmann-odd
                one-form on $\Sigma$ with values in the Lie algebra $\lieg$, and the action is schematically modified as
		\begin{equation*}
			S_{\textrm{\tiny YM}} \longmapsto
                        S_{\textrm{\tiny YM}}  + \int_{\Sigma}\, \tr(\psi
                        \wedge \psi) \ .
		\end{equation*}
		The equivalence is straightforwardly checked by
                integrating out $\psi$. On the other hand, the new
                term is topological and hence is insensitive to the $T
                \overline{T}$-deformation. We can thus first $T
                \overline{T}$-deform and integrate out $\psi$
                afterwards, obtaining again the result
                \eqref{eq:ZqYMTT}. So regardless of the route
                followed, the $T \overline{T}$-deformation of
                two-dimensional Yang-Mills theory and its
                generalizations always provides the same answer with
                or without supersymmetry.

                \subsubsection*{Refinement}
                
		Let us now consider the refinement of $q$-deformed
                Yang-Mills theory~\cite{Aganagic2012}, also known as
                $(q,t)$-deformed Yang-Mills theory. The refinement
                leaves the action unchanged but modifies the path
                integral measure~\cite{Aganagic2012}. Therefore we can
                $T \overline{T}$-deform the theory and the
                abelianization technique continues to work, hence the $T
                \overline{T}$-deformation modifies the partition
                function of $(q,t)$-Yang-Mills theory only in the
                Gaussian potential, according to \eqref{eq:Vttdef}. We will
                give more details later on in Section \ref{sec:tqTTLargeN}.

                \subsubsection*{$\boldsymbol\theta$-angle}
                
		In $q$-Yang-Mills theory, the $\theta$-angle term is
                introduced in the path integral as a linear term in
                the potential $V (\phi)$, and it descends from a
                chemical potential for D2-branes in the construction
                of~\cite{Vafa2004,Aganagic2004}. Therefore it will
                also couple to the gravitational path integral after
                $T \overline{T}$-deformation, and will enter in the
                denominator of the deformed potential through
		\begin{equation*}
			- \frac{\lambda}{2N}\, C_2 (R) + \I\, \theta\,
                        C_1 (R) \longmapsto	\frac{   -
                          \frac{\lambda}{2N}\, C_2 (R) + \I\, \theta\,
                          C_1 (R) }{ 1 - \frac{\tau}{N^3} \left[ C_2
                            (R) - \frac{ 2\,\I\, \theta\,
                              N}{\lambda}\, C_1 (R) \right] }  \ ,
		\end{equation*}
		where $C_1$ is the first Casimir of $G$, which is non-zero only for non-simply connected gauge groups.

                \subsection{Wilson loops, marked points and $q$ a root of unity}
		
		One may also include Wilson loop operators in an
                irreducible representation $R$ of $G$ along a closed
                curve $\mathcal{C}$ on $\Sigma$:
		\begin{equation*}
			W_R ( \mathcal{C}) = \tr _R\, \mathcal{P} \exp
                        \oint_{\mathcal{C}}\, A \ .
		\end{equation*}
		We assume for simplicity that $\mathcal C$ does not
                wind around any handle of $\Sigma$.
                
\begin{figure}[htb]
		\centering
		\includegraphics[width=0.2\textwidth]{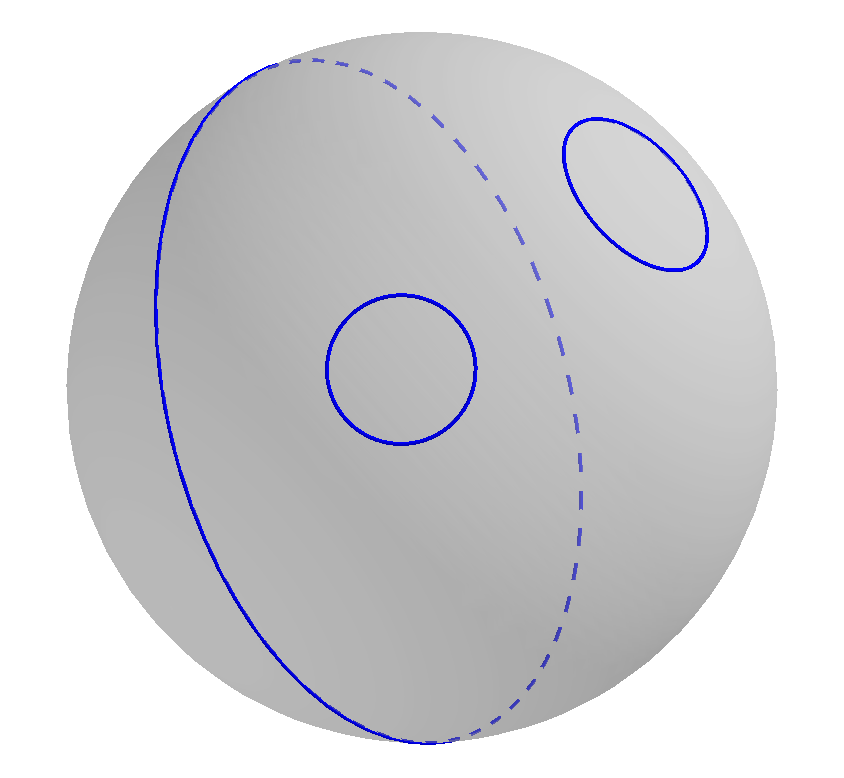}%
		\hspace{0.1\textwidth}%
			\begin{tikzpicture}[tqft/flow=east]
			\begin{scope}[tqft/boundary upper style={draw,blue,thick},tqft/boundary lower style={draw,blue,thick}]
			\node[tqft/pair of pants,draw,fill=gray,fill opacity=0.5] (a) at (0,0) {};
			\node[tqft/cap,draw,fill=gray,fill opacity=0.5] (a) at (-3,0) {};
			\node[tqft/cup,draw,fill=gray,fill opacity=0.5] (a) at (3,1) {};
			\node[tqft/cup,draw,fill=gray,fill opacity=0.5] (a) at (3,-1) {};
			\end{scope}
			\end{tikzpicture}
		\caption{\small A sphere with three Wilson loops is cut into
                  three disks plus a remaining pair of pants.}
		\label{fig:WLcut}
		\end{figure}
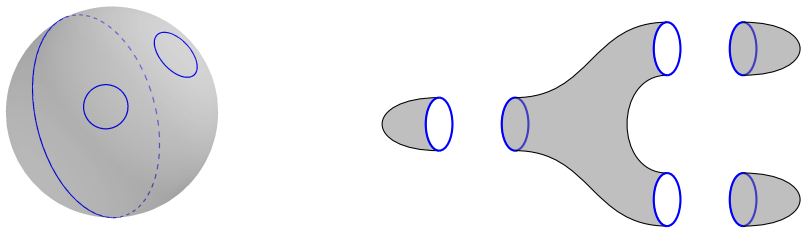

		The expectation value of a collection of $s$ Wilson
                loops in $T \overline{T}$-deformed two-dimensional
                Yang-Mills theory, both the ordinary and $q$-deformed
                versions, is computed as follows. Cut $\Sigma$ along
                the $s$ cycles $\mathcal{C}_1, \dots, \mathcal{C}_s $,
                obtaining $s+1$ components: $s$ of them have disk
                topology and the last is the remainder of Euler
                characteristic $\chi (\Sigma) - s$. The next step is
                to compute the $T \overline{T}$-deformed partition
                function on each component, which is a wavefunction of
                the holonomies along the boundaries $\mathcal{C}_1,
                \dots ,\mathcal{C}_s$. Then glue the components
                pairwise back together by integrating over $G$. The
                example of $\mathbb{S}^2$ with three loops is depicted
                in Figure~\ref{fig:WLcut}.
                
		In this way we find the normalized expectation value
		\begin{align*}
                                                            \big\langle
                                                            W_{R_1}(\mathcal{C}_1)\cdots
                                                                   W_{R_s}(\mathcal{C}_s)\big\rangle
                  & = \frac{1}{ \mz_{\textrm{\tiny $q$-YM}} ^{T
                    \overline{T}}[\Sigma] } \, \sum_{R} \,  \dim_q
                    (R)^{\chi (\Sigma) - s} \, q^{\frac{p - A_W}{2}\, C_2 ^{T \overline{T}} (R,\tau)   } \\ 
			& \hspace{3cm} \times \ \prod_{i=1} ^{s} \
                   \sum_{\widetilde{R}_i}\, \dim_q (\widetilde{R}_i) \,
                   q^{\frac{a_i}{2}\, C_2 ^{T \overline{T}}
                   (\widetilde{R}_i,\tau) } \, { N^{\widetilde{R}_i } }
                   ^{\phantom{\dag}}_{ R_i  R} \ ,
		\end{align*}
		where $a_i$ is the area enclosed by the loop
                $\mathcal{C}_i$, $R_i$ is the representation label of
                the $i$-th loop, and $\widetilde{R}_i$ is a summation
                variable denoting an irreducible representation
                associated to the quantization on the $i$-th
                component. Geometrically, $R$ is associated to the
                remainder, $R_i$ to the $i$-th cut and
                $\widetilde{R}_i$ to the $i$-th disk. We have also
                denoted
                $$
                A_W = \sum_{i=1}^s\, a_i \ ,
                $$
                and we assume $A_W<p$ to ensure convergence of the
                first series. The quantities ${N^{R_1}}^{\phantom{\dag}} _{R_2 R_3}$ are
                fusion coefficients obtained from the integration over
                the holonomies; for unitary gauge group they are the
                Littlewood-Richardson coefficients. This formula
                differs from the original theory simply in the
                replacement of the Casimir as in \eqref{eq:C2toTT}.
                
		In the limit in which the loops shrink to a point,
                $a_i\to0$, we obtain (after dropping the normalization
                $\mathcal{Z}_{\textrm{\tiny
                    $q$-YM}}^{T\overline{T}}[\Sigma]^{-1}$) the
                partition function of $T \overline{T}$-deformed
                $q$-Yang-Mills theory on a surface $\Sigma$ with $s$ marked points decorated with irreducible representations $R_{1}, \dots , R_{s}$:
		\begin{equation*}
			\mz_{\textrm{\tiny $q$-YM}} ^{T \overline{T}}
                        [ \Sigma ; R_1, \dots , R_s ] = \sum_R \,
                        \dim_q (R)^{\chi (\Sigma) }  \,
                        q^{\frac{p}{2}\, C_2 ^{T \overline{T}} (R,\tau) } \
                        \prod_{i=1} ^{s} \ \sum_{\widetilde{R}_i} \, \frac{ \dim_q ( \widetilde{R}_i ) }{ \dim_q (R) } \,  { N^{ \widetilde{R}_i } }^{\phantom{\dag}} _{R_i R} \ .
                      \end{equation*}
                      
		Analogous formulas hold for the $T
                \overline{T}$-deformation of ordinary Yang-Mills theory.
		
		\subsubsection*{Lost connection with Chern-Simons
                  theory}
                
		For $\tau=0$ and $0<\vert q \vert <1$ (with possibly
                $q \in \mathbb{C}$), when $\vert q \vert \to 1$ at
                roots of unity, the sum over representations
                terminates for gauge group $G=U(N)$
                \cite{naculich2007level}. After $T
                \overline{T}$-deformation, the quadratic Casimir part
                is modified and the cancellations that truncate the
                series no longer take place.
                
		The usual connection between $q$-Yang-Mills theory and
                Chern-Simons theory thus no longer holds. For the same
                reason, even when $q$ is a root of unity, one cannot
                understand Wilson loops in the $T
                \overline{T}$-deformed version in terms of observables
                in Chern-Simons theory, or some deformation thereof,
                living in the total space of a degree~$p$ circle
                bundle over $\Sigma$.
                
		Furthermore, when looking for the modular matrices
                $\mathsf{S}$ and $\mathsf{T}$ of $PSL (2 ,
                \mathbb{Z})$ in expressions such as \eqref{eq:ZqYMTT}, we recall that matrix elements like 
		\begin{equation*}
			\mathsf{S}_{R \widetilde{R}} \qquad \text{and} \qquad \mathsf{T}_{R \widetilde{R}}
		\end{equation*}
		are defined for integrable representations $R$ and $
                \widetilde{R}$, while in our case the sum runs over
                all the irreducible representations. In this sense the
                $T \overline{T}$-deformation spoils the modularity
                properties of the theory at $q$ a root of unity, as
                could have been foreseen from the explicit form of  \eqref{eq:C2toTT}.\footnote{Gauging the $T \overline{T}$-deformed WZW model of \cite{Bonelli2018} does not yield a connection with Chern-Simons theory, or some deformation thereof. The relation with Chern-Simons theory is indeed a special property of the conformal fixed point~\cite{MooreSeiberg}.}

	\subsection{Breakdown of factorization}
	
	We have seen that the usual connection with Chern-Simons
        theory is lost as soon as the $T \overline{T}$-deformation is
        turned on. In the following we discuss another well-known
        central feature of two-dimensional Yang-Mills theory that does
        not hold after $T \overline{T}$-deformation: the factorization
        of the partition function $\mz_{\textrm{\tiny $q$-YM}}$ into
        chiral and anti-chiral sectors
        \cite{Vafa2004,Aganagic2004,Caporaso2005,Caporaso2005chiral}. This
        strongly suggests that the usual
        large $N$ string theory picture of two-dimensional Yang-Mills
        theory breaks down after $T\overline{T}$-deformation.

        Let $\mathscr{H}$ be the Hilbert space of states of the
        theory, and endow it with the basis $\left\{ \vert R \rangle
        \right\}$ in one-to-one correspondence with isomorphism
        classes of irreducible unitary representations of $G$~\cite{CMRlectures}. Adopting a common shorthand, we call it the representation basis. The normalization is 
	\begin{equation*}
		\langle R \vert \widetilde{R}\, \rangle = \dim( R
                )^{\chi (\Sigma)} \ \delta_{R \widetilde{R}} \ .
              \end{equation*}
              
	The factorization property relies on being able to make a replacement
	\begin{equation*}
		q^{\frac{p}{2}\, C_2 (R) } \longrightarrow
                q^{\frac{p}{2}\, C_2 (R_+) } \, q^{\frac{p}{2}\, C_2
                  (R_- ) }  \ ,
	\end{equation*}
	where $R_{+}$ and $R_-$ are known as ``chiral'' and ``anti-chiral'' representations, which correspond to states 
	\begin{equation*}
		\vert R_{\pm} \rangle \in \mathscr{H}_{\pm} \ ,
	\end{equation*}
	in the factorized Hilbert space $\mathscr{H}_+ \otimes \mathscr{H}_-$. It is clear from \eqref{eq:C2toTT}--\eqref{eq:ZqYMTT} that the factorization breaks down at $\tau \ne 0$.
	
	\subsubsection*{Quantization of the $\boldsymbol{T \overline{T}}$-deformed
          theory}
        
	Consider the unitary gauge group $G= U(N)$ and the surface
        $\Sigma$ as a fibration over $\mathbb{S}^1$, with the circle
        interpreted as the Euclidean time direction. By definition,
        the partition function of $q$-Yang-Mills theory on $\Sigma$ is
        given by
	\begin{equation*}
		\mathcal{Z}_{\textrm{\tiny $q$-YM}} (\lambda ) =
                \mathrm{Tr} _{\mathscr{H}}\, \e^{-
                  \frac{\lambda\,p}{2N}\, \widehat{H}_{\textrm{\tiny
                      $q$-YM}}} = \sum_R \, \langle R \lvert \e^{-
                  \frac{\lambda\,p}{2N}\, \widehat{H}_{\textrm{\tiny
                      $q$-YM}}} \rvert R \rangle  \ ,
	\end{equation*}
	where $ \widehat{H}_{\textrm{\tiny $q$-YM}}$ is the
        Hamiltonian, and we have taken the trace over the Hilbert
        space $\mathscr{H}$ in the representation basis, which
        diagonalizes $ \widehat{H}_{\textrm{\tiny $q$-YM}}$ with
        eigenvalues $C_2(R)$. A generic
        deformation controlled by a parameter $\tau$ which triggers an RG flow would produce 
	\begin{equation*}
		\mathcal{Z}_{\textrm{\tiny $q$-YM}} ^{\text{def}}
                (\lambda, \tau ) = \mathrm{Tr} _{\mathscr{H} (\tau)}\,
                \e^{- \frac{\lambda\,p}{2N}\, \widehat{H} (\tau) } =
                \sum_{R (\tau)}\, \langle R (\tau)  \lvert \e^{-
                  \frac{\lambda\,p}{2N}\, \widehat{H} (\tau) } \rvert R
                (\tau) \rangle \ ,
	\end{equation*}
	deforming both the Hamiltonian to $\widehat{H}(\tau)$ and the
        Hilbert space to $\mathscr{H}(\tau)$. The basis $ \left\{
          \vert R (\tau) \rangle \right\} $ would reduce to the
        representation basis when sending $\tau \to 0$. Note that, in
        this general framework, since the Hilbert space changes, one
        may need to include additional states. However, when the
        deformation is by the composite operator $T\, \overline{T}$,
        the explicit form of the deformed eigenvalues of $\widehat{H}
        (\tau)$ is known, and in particular no new eigenvalues
        arise. For $q$-Yang-Mills theory we obtain explicitly 
	\begin{equation*}
		\mathcal{Z}_{\textrm{\tiny $q$-YM}} ^{T \overline{T}}
                (\lambda, \tau )  = \sum_{R (0)}\, \langle R (0)
                \lvert \e^{- \frac{\lambda\,p}{2N}\, \widehat{H} (\tau) }
                \rvert R (0) \rangle \ ,
	\end{equation*}
	with the deformed Hamiltonian
        \begin{align*}
\widehat{H} (\tau) = \frac{\widehat{H}_{\textrm{\tiny
          $q$-YM}}}{1-\frac\tau{N^3}\, \widehat{H}_{\textrm{\tiny $q$-YM}}}
        \end{align*}
        diagonalized by the
        representation basis $ \left\{ \vert R \rangle \right\} =
        \left\{ \vert R (0) \rangle \right\} $ for all $\tau \ge 0
        $. Only the eigenvalues \smash{$C_2^{T\overline{T}}(R,\tau)$} are different. Therefore, although in
        general from the knowledge of the eigenvalues one cannot
        exclude that additional degenerate states arise in the $T
        \overline{T}$-deformed theory, we see that this is not the
        case for two-dimensional Yang-Mills theory and its
        relatives. Indeed, having found explicitly the $T
        \overline{T}$-deformed partition function, the presence of
        additional states at $\tau >0$ should have a null net
        contribution, but this is not possible from the explicit, strictly positive form
        of the eigenvalues.
        
	In conclusion, the partition function of $T
        \overline{T}$-deformed $q$-Yang-Mills theory, and hence also
        ordinary Yang-Mills theory through the limit
        \eqref{eq:qtoYMlimit}, is the trace of the exponential of the $T
        \overline{T}$-deformed Hamiltonian found in \cite{Conti2018}
        taken in an undeformed Hilbert space of states.
	
	\subsubsection*{Free fermion formulation}
        
		Let us now focus on ordinary (without $q$-deformation)
                two-dimensional Yang-Mills theory for
                definiteness. While the factorization structure of
                $q$-Yang-Mills theory is richer, in the sense that
                already at finite $N$ one sees a factorization into
                chiral and anti-chiral building blocks, the breakdown
                of these properties happens at a fundamental level,
                which is more clearly seen by looking directly at
                $T\overline{T}$-deformed ordinary Yang-Mills theory.
                
		It is well-known that the Hilbert space $\mathscr{H}$
                factorizes at large $N$ as \cite{GrossTaylor} 
		\begin{equation}
		\label{eq:Hilbertfactor}
			\mathscr{H} \xrightarrow{ \ N \to \infty \ } \mathscr{H}_+ \otimes \mathscr{H}_- \ ,
		\end{equation}
		with the representation basis factorizing accordingly as
		\begin{equation*}
			\vert R \rangle  \xrightarrow{ \ N \to \infty \ }  \vert R_+ \rangle \otimes \vert R_{-} \rangle \ ,
		\end{equation*}
		with $ \vert R_{\pm} \rangle \in
                \mathscr{H}_{\pm}$. The Hilbert spaces
                $\mathscr{H}_{+} $ and $\mathscr{H}_-$ are known as
                the ``chiral'' and ``anti-chiral'' sectors, respectively. From the factorization \eqref{eq:Hilbertfactor} one has \cite{GrossTaylor} 
		\begin{align*}
			\lim_{N\to\infty}\, \mathcal{Z}_{\textrm{\tiny YM}}(A)
                  &= \bigg(
                                                     \sum_{R_+}\,
                                                     \langle R_+ \vert
                                                     \e^{-
                                                     \frac{A}{2N}\,
                                                     \widehat{H}_{\textrm{\tiny
                                                     YM}} } \vert R_+
                                                     \rangle
                                                     \bigg)\,\bigg(
                                                     \sum_{R_-}\,
                                                     \langle R_- \vert
                                                     \e^{-
                                                     \frac{A}{2N}\,
                                                     \widehat{H}_{\textrm{\tiny
                                                     YM}}  } \vert R_- \rangle \bigg) \\[4pt]
				& = \left( \mathrm{Tr}_{
                           \mathscr{H}_+}\, \e^{- \frac{A}{2N}\,
                           \widehat{H}_{\textrm{\tiny YM}}  }
                           \right) \left( \mathrm{Tr}_{
                           \mathscr{H}_-}\, \e^{- \frac{A}{2N}\,
                           \widehat{H}_{\textrm{\tiny YM}}  }   \right) 
		\end{align*}
                where we dropped overall constants.
		In the $T \overline{T}$-deformed theory, the large $N$
                factorization of the Hilbert space
                \eqref{eq:Hilbertfactor} continues to hold according to the
                discussion above, but the trace can no longer be
                factorized into a product of traces.
                
\begin{figure}[htb]
				\centering
				\includegraphics[width=0.2\textwidth]{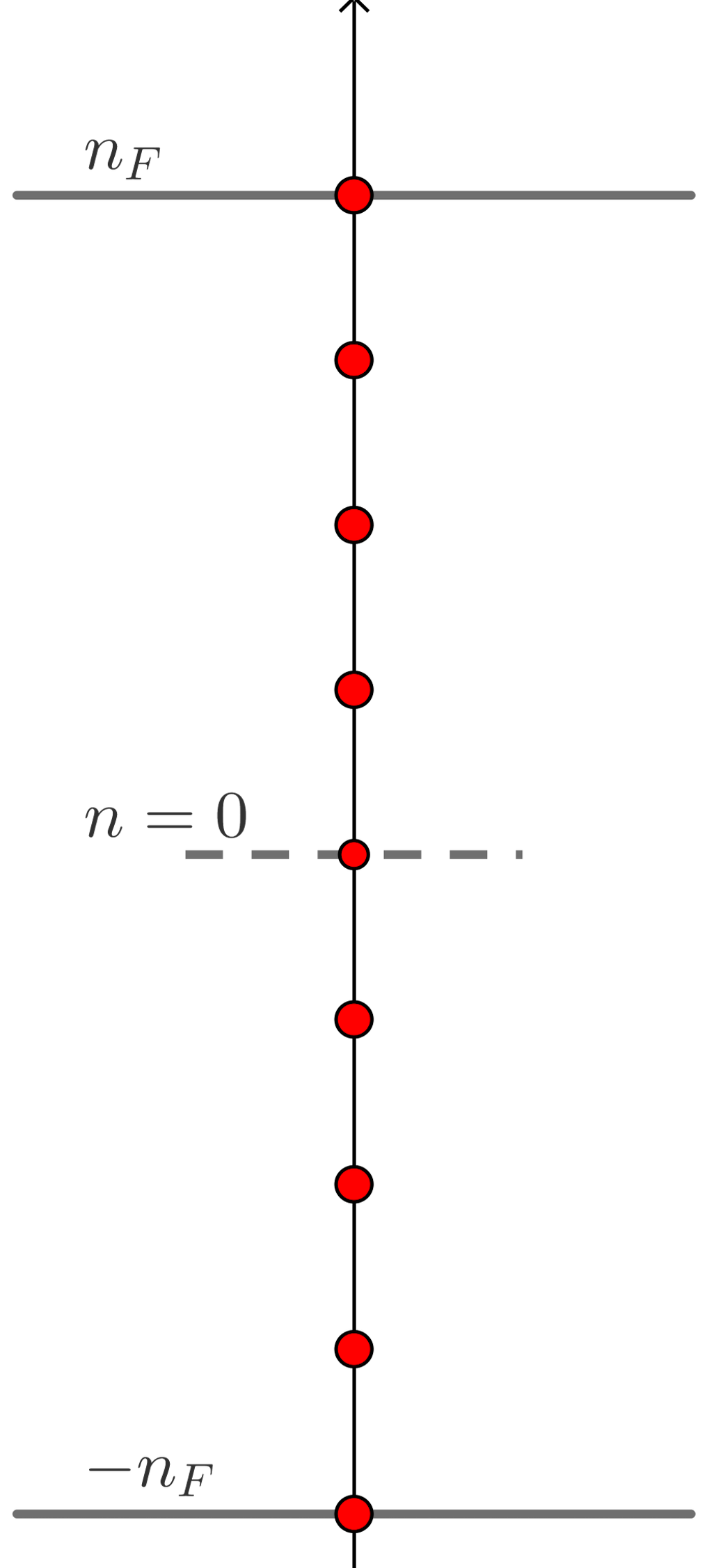}%
				\hspace{0.1\textwidth}
				\includegraphics[width=0.2\textwidth]{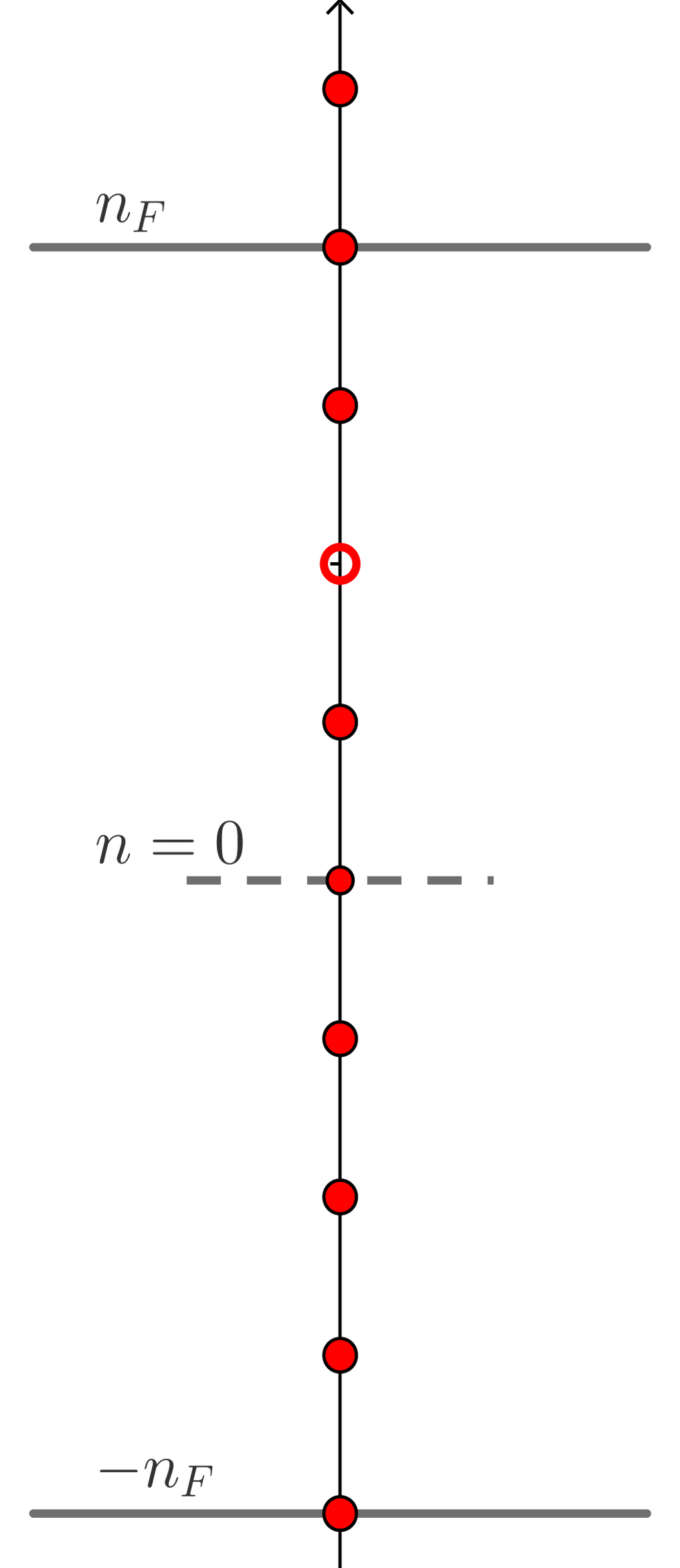}%
				\hspace{0.1\textwidth}
				\includegraphics[width=0.2\textwidth]{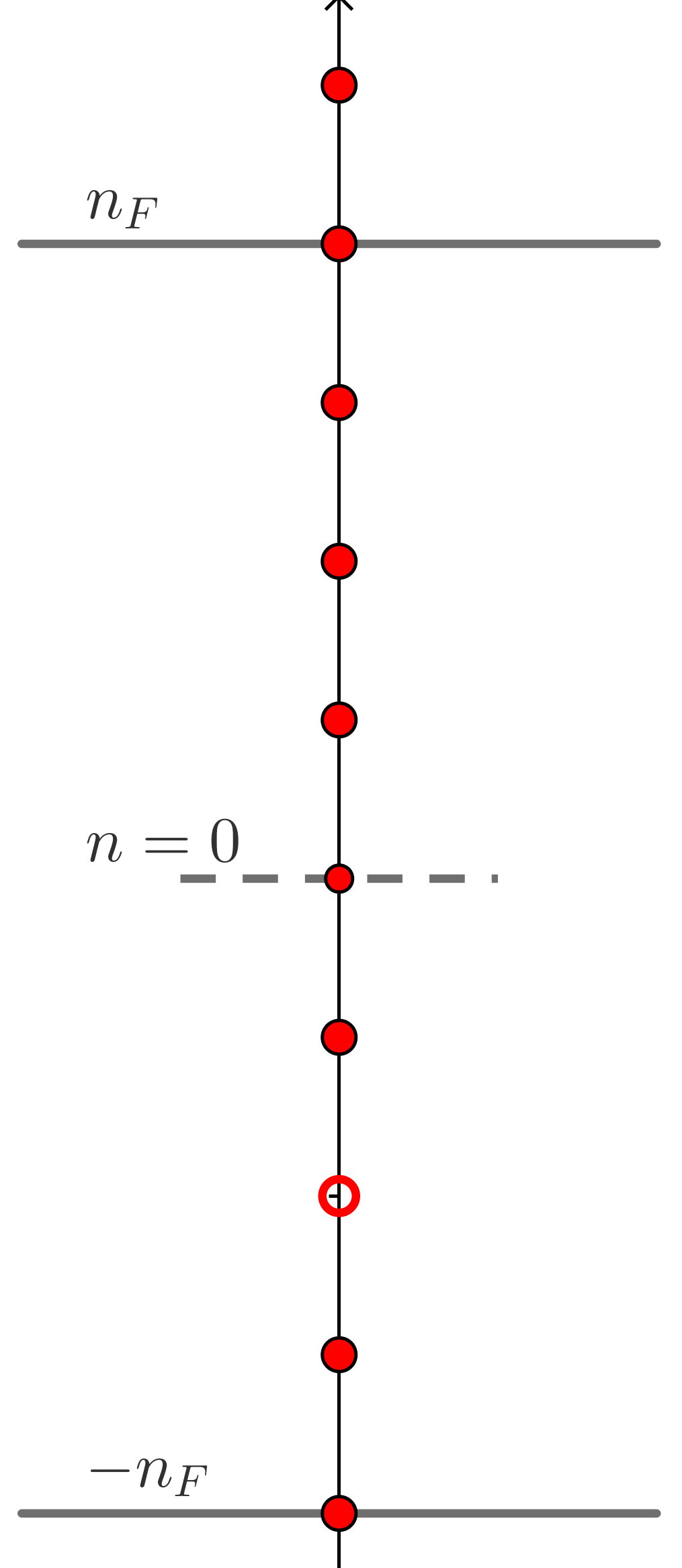}
				\caption{\small Non-relativistic
                                  fermions: ground state (left) and
                                  two excited states (center,
                                  right). The two excited states are
                                  excitations over the positive Fermi
                                  surface. In the center, a fermion
                                  occupying a positive energy level
                                  jumps above the positive Fermi
                                  surface: this will correspond to a
                                  chiral state at large $N$. On the
                                  right, a fermion occupying a
                                  negative energy level jumps above the positive Fermi surface: this will be exponentially suppressed at large $N$.}
			\label{fig:fermionsE}
			\end{figure}

		We will now further elucidate this point through
                the equivalence with a system of $N$ non-relativistic
                fermions \cite{Douglas1993Group,Douglas1993YM}
                (non-perturbative corrections were studied in
                \cite{DGOV2005}). In the mapping of two-dimensional $U(N)$
                Yang-Mills theory to a system of $N$ free fermions on
                $\mathbb{S}^1$ \cite{Douglas1993Group,Douglas1993YM},
                the ground state corresponds to the state where the
                fermions occupy the $N$ lowest energy levels, as in
                the left panel of Figure \ref{fig:fermionsE}. In the
                representation basis, the ground state is described by
                the trivial representation, while higher-dimensional
                representations are mapped to excited states, in which
                fermions have jumped to higher energy levels.
		
		\par
		
		At finite $N$, excitations above the positive Fermi
                surface, or below the negative Fermi surface, may
                arise from any fermion, as depicted in the central and
                right panels of Figure \ref{fig:fermionsE}. In the
                large $N$ limit the two Fermi surfaces decouple, and
                excitations above the positive Fermi level
                (respectively below the negative Fermi level)
                correspond to fermions close to that surface, thus
                with positive (respectively negative) energy, jumping
                to higher (respectively lower) unoccupied levels.
                
		Jumps of order $N$ sites are exponentially suppressed
                with $N$, and can only be seen from a non-perturbative analysis \cite{DGOV2005}. Therefore the factorization is interpreted as a disentanglement of the Fermi surfaces.
			
			In the $T \overline{T}$-deformed theory the
                        two Fermi surfaces remain entangled even in
                        the large $N$ limit. The crucial difference
                        between the picture of \cite{Douglas1993Group}
                        and its $T \overline{T}$-deformed version lays
                        in the interpretation of the Casimir in terms
                        of free fermions. While at $\tau =0$ it is a
                        confining quadratic potential, this
                        interpretation is lost at $\tau>0$. Indeed, by
                        expanding the $T \overline{T}$-deformed
                        potential \eqref{eq:Vttdef} in a geometric
                        series, we do not obtain a confining potential
                        for fermions, but instead infinitely many
                        non-local interaction terms which introduce
                        long-distance correlations.
                        
			A consequence of these additional interactions
                        is that, even at $N \to \infty$, the energy
                        required for a fermion to jump to another
                        level does not only depend on the energy
                        separation between the initial and final
                        state, but it is also a function of the levels
                        occupied by all of the other fermions. From a
                        conformal field theory perspective, this casts
                        doubt on the existence of a string theory dual
                        to $T\overline{T}$-deformed two-dimensional
                        Yang-Mills theory, unless it is a highly
                        exotic one.

                        \section{Phase transitions in $\boldsymbol{T
                            \overline{T}}$-deformed $\boldsymbol q$-Yang-Mills theory}
	\label{sec:phaseTTqYM}
	
\begin{figure}[htb]
		\centering
		\includegraphics[width=0.5\textwidth]{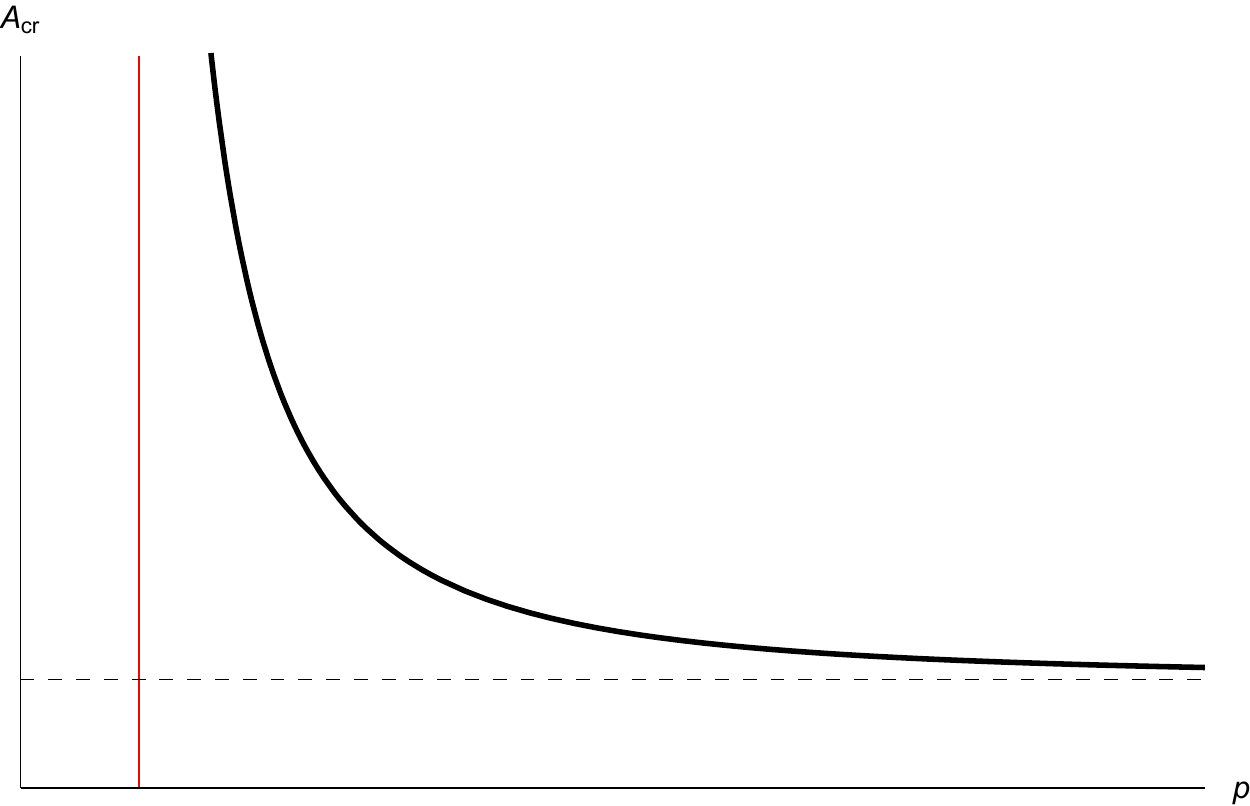}
		\caption{\small The critical curve of $q$-Yang-Mills
                  theory, in terms of the parameter $A= \lambda/p$ as
                  a function of $p$. The horizontal asymptote (dashed)
                  is the DK critical point $A_{\text{cr}}= \pi^2$. The
                  vertical asymptote (red) is the point $p=2$. This plot is inspired by \cite{Arsiwalla}.}
		\label{fig:crit-qdef}
		\end{figure}

		Two-dimensional $U(N)$ Yang-Mills theory on
                $\mathbb{S}^2$ undergoes a third order phase
                transition \cite{DK93}, henceforth called the
                Douglas-Kazakov (DK) transition, which is induced by
                instanton instabilities~\cite{Gross1994}. The same
                third order phase transition is experienced by the
                $q$-deformed theory
                \cite{Arsiwalla,JafferisMarsano,Caporaso2005} when
                $p>2$. The critical value of the coupling
                $\lambda_{\text{cr}}$ decreases monotonically with
                increasing $p$ and one eventually recovers the DK
                transition in the limit $p \to \infty$
                \cite{Arsiwalla,JafferisMarsano,Caporaso2005}, see
                Figure~\ref{fig:crit-qdef}. This means that the
                $q$-deformation extends the region in parameter space
                corresponding to a weak coupling phase.
                
		In \cite{Santilli2018} it was shown that $T
                \overline{T}$-deformed (but not $q$-deformed)
                Yang-Mills theory also undergoes a DK-type transition for $0 \le \tau < \tau_{\text{max}}$, with 
		\begin{equation}
		\label{eq:taumax}
			\frac{1}{\tau_{\text{max}} } = \frac{1}{\pi^2}
                        - \frac{1}{12} \ .
		\end{equation}
		The critical value of the area parameter decreases
                with increasing $\tau$, and eventually no weak coupling phase exists when $\tau$ approaches $\tau_{\text{max}}$ \cite{Santilli2018}. Therefore the $T \overline{T}$-deformation reduces the region of parameter space corresponding to a weak coupling phase.\par
		
		The goal of this section is to analyze the large $N$
                phase structure when both the $q$-deformation and the
                $T \overline{T}$-deformation are turned on. In the
                following we summarize our large $N$ results. Then we
                proceed in Section \ref{sec:reviewTT1} to briefly
                recall the strategy of \cite{Santilli2018} to extend
                the analysis of \cite{DK93,Gross1994} to the $T
                \overline{T}$-deformed setting. In Section
                \ref{sec:TTphase1} we present the large $N$ formalism
                when both deformations are turned on and study the
                weak coupling regime, while Section \ref{eq:critical}
                is dedicated to a study of the critical
                surface. Section \ref{sec:instantons} discusses the
                role of instanton contributions, while Section
                \ref{sec:TTphase2} is dedicated to a study of the phase transition and the strong coupling regime. Finally, in Section \ref{sec:tqTTLargeN} we comment on the large $N$ limit of the refined theory.
		
		\subsubsection*{Large $\boldsymbol N$ results}
                
			Before diving into the detailed analysis of
                        the large $N$ phase structure, we summarize here our main findings:
			\begin{itemize}
				\item $T \overline{T}$-deformed
                                  $q$-Yang-Mills theory undergoes a third order phase transition when $p>p_0$. Remarkably, we find that $p_0 <2$.
				\item The slice of parameter space
                                  giving a weak coupling phase is
                                  extended, relative to the pure
                                  $T \overline{T}$-deformation, and is
                                  reduced only relative to the $q$-deformation. This interpolates perfectly between the effects discovered respectively in \cite{Arsiwalla,JafferisMarsano,Caporaso2005} and \cite{Santilli2018}, recovering the single-deformation scenarios as limiting cases.
				\item The phase transition is induced
                                  by instantons. The shrinking of the
                                  weak coupling region is explained by
                                  the fact that in the $T \overline{T}$-deformed theory the suppression factor of the instantons is smaller, hence their effect becomes relevant at lower values of the coupling $\lambda$.
			\end{itemize}
		
		\subsection{Large $N$ limit of $T
                  \overline{T}$-deformed Yang-Mills theory}
		\label{sec:reviewTT1}
                
		Here and in the rest of this section we take $G=U(N)$ and $\Sigma=\mathbb{S}^2$. We also continue the Euler characteristic $\chi$ to real values close to $\chi (\mathbb{S}^2)=2$, 
		\begin{equation}
		\label{eq:continuechi}
			0 < 2 - \varepsilon_- < \chi < 2 +
                        \varepsilon_+ \ ,
		\end{equation}
with $\varepsilon_\pm>0$,		which will also be useful
later on in Section~\ref{sec:entanglement}.

		Irreducible $SU(N)$ representations up to isomorphism
                are in one-to-one correspondence with Young diagrams,
                labelled by partitions $\mathfrak{R}= (\mathfrak{R}_1, \dots, \mathfrak{R}_{N-1})$ with 
		\begin{equation*}
			\mathfrak{R}_1 \ge \mathfrak{R}_2 \ge \cdots
                        \ge \mathfrak{R}_{N-1} \ge \mathfrak{R}_{N} :=
                        0 \ .
		\end{equation*}
		Using the short exact sequence of groups
		\begin{equation*}
			1 \longrightarrow SU(N) \longrightarrow U(N) \longrightarrow U(1) \longrightarrow 1 
		\end{equation*}
		the irreducible $U(N)$ representations are obtained
                from those of $SU(N)$ summing over the $U(1)$
                sector. Therefore a $U(N)$ representation $R$
                corresponds to a partition
		\begin{equation}
			\label{eq:RWeyl}
				+ \infty > R_1 \ge R_2 \ge \cdots \ge
                                R_N > - \infty \ ,
		\end{equation}
		with $R_i = \mathfrak{R}_i + \mathfrak{r}$ for all $i=1, \dots, N$ and $\mathfrak{r} \in \mathbb{Z}$. It is useful to change variables 
		\begin{equation}
		\label{eq:changeh}
			h_i = - R_i + i - \frac{N+1}{2} \qquad
                        \mbox{for} \quad i =1, \dots, N \ .
                      \end{equation}
                      
		In these variables the partition function of
                $T\overline{T}$-deformed $U(N)$ Yang-Mills theory reads 
		\begin{equation}
		\label{eq:ZYMTTh}
			\mz_{\textrm{\tiny YM}} ^{T \overline{T}}(A,\tau)
                        = \frac{1}{N!\, G(N+1)^{\chi}} \, \sum_{\vec h \in
                          \mathbb{Z}^{N} }\,  \Delta (\vec h)^{\chi} ~ \exp
                        \Bigg( - \frac{ \frac{ A }{2 N} \, \Big(
                          \sum\limits_{j=1 } ^{N}\, h_j ^2 - \frac{
                            N\, (N^2 +1)}{12}  \Big) }{ 1 - \frac{
                            \tau}{N^3}\, \Big( \sum\limits_{j=1 }
                          ^{N}\, h_j ^2 - \frac{ N\, (N^2 +1)}{12}
                          \Big) }  \Bigg) \ ,
		\end{equation}
		where we used the symmetry of the sum to lift the
                restriction \eqref{eq:RWeyl} to the principal Weyl
                chamber, letting the sum run over unordered
                $\vec h=(h_1,\dots,h_N) \in \mathbb{Z}^{N}$. The
                shift proportional to $-\frac{1}{12}$ in the Casimir
                would simply give an overall factor at $\tau=0$, but
                it becomes relevant at $\tau>0$. Here $G$ is the Barnes $G$-function, which for integer argument can be written as 
		\begin{equation}
		\label{eq:BarnesG}
			G(N+1) = \prod_{j=1} ^{N-1}\, j! 
		\end{equation}
		and $\Delta (\vec h)$ is the Vandermonde determinant 
		\begin{equation}\label{eq:Vandermonde}
			\Delta (\vec h) = \prod_{1 \le i < j \le N }\, (h_i
                        - h_j ) \ .
		\end{equation}

		When sending $N \to \infty $ the leading contribution to the partition function comes from the saddle point configuration. This is the one that minimizes the action 
		\begin{equation*}
			S_{\textrm{\tiny YM}} [h] = \frac{A}{2}\, \sum_{k=0} ^{\infty}\,
                        \tau^k \left( \int_0 ^{1} h(x)^2\, \dd x -
                          \frac{1}{12}  \right)^{k+1} -
                        \frac{\chi}{2}\, \int_0 ^{1}\, \dd x \ \int_0
                        ^{1}\, \dd y \, \log \vert h(x)- h(y) \vert \ ,
		\end{equation*}
		where $x= i/N$ and $h(x)= h_i /N$ have become
                continuous variables at large $N$, and we have
                expanded the $T \overline{T}$-deformed Casimir in a
                geometric series, which is allowed as long as the $T
                \overline{T}$-deformed theory is well-defined. We are
                also taking a 't~Hooft limit, since we already
                reabsorbed the gauge coupling in the definition of
                $A/N$.
                
		Introduce
                the eigenvalue density $\rho (h)$, which is defined by 
		\begin{equation*}
			\rho (h)\, \dd h = \dd x 
		\end{equation*}
		and is normalized 
		\begin{equation*}
			\int_{\text{supp}(\rho)} \, \rho (h)\, \dd h =1 \ .
		\end{equation*}
	Then taking the derivative of the action and setting it equal
        to zero, we arrive at the saddle point equation 
		\begin{equation}
		\label{eq:SPETT}
			\ncint\ \dd u \, \frac{\rho (u)}{h-u} =
                        \frac{ A}{\chi}\, h\, \sum_{k=0}^{\infty}\,
                        (k+1)\, \tau^k \, \left( \mu_2 - \frac{1}{12}
                        \right)^{k} \ ,
		\end{equation}
		where $\ncint$ denotes the principal value of the
                integral over the support and $\mu_2$ is the second moment of the eigenvalue distribution $\rho$:
		\begin{equation}
		\label{eq:defmu2}
			\mu_2 := \mu_2 [\rho] =
                        \int_{\text{supp}(\rho)} \, \dd u\, \rho (u)\,
                        u^2 \ .
		\end{equation}
		A crucial aspect here is that the original ensemble is
                discrete, and the condition \eqref{eq:RWeyl} in terms
                of the new weight variables $h_i$ says that 
		\begin{equation*}
			h_{i+1} - h_i \ge 1 \ , 
		\end{equation*}
		which in the large $N$ limit implies 
		\begin{equation}
                  \label{eq:condrhole1}
			\rho (h) \le 1 \qquad \mbox{for} \quad h \in
                        \text{supp}( \rho ) \ . 
		\end{equation}
		The solutions we find will have to satisfy this
                constraint: in a generic discrete ensemble, if
                $\rho(h)$ ceases to satisfy \eqref{eq:condrhole1} at a
                codimension one locus in the space of parameters, the
                system undergoes a phase transition, and the new phase
                will be governed by a different density $\rho (h)$.
		
		\subsubsection*{Strategy}
                
			Standard methods for solving integral equations
                        \cite{Pipkin} do not apply to the saddle point
                        equation \eqref{eq:SPETT} due to the
                        dependence on $\rho$ on both sides of the
                        equality. However, it was noted in
                        \cite{Santilli2018} that \eqref{eq:SPETT} can
                        be solved perturbatively in $\tau$, with the
                        zeroth order being the DK solution
                        \cite{DK93}.
                        
			The technique of \cite{Santilli2018}
                        essentially consists in solving the equation
                        at $k$-th order using the expression for
                        $\mu_2$ obtained by plugging in the distribution
                        $\rho (h)$ approximated at $(k-1)$-th order,
                        which is the standard method of iteration in
                        perturbation theory. The solution can be found
                        in this way thanks to the especially simple
                        dependence of the right-hand side of
                        \eqref{eq:SPETT} on $\rho (h)$: it only enters
                        the saddle point equation in a coefficient
                        that renormalizes $A$. In fact, we can write
                        the saddle point equation at a generic order $O (\tau^{k})$ as 
			\begin{equation*}
				 \ncint\ \dd u \, \frac{\rho (u)}{h-u}
                                 = \frac{ A}{\chi}\, c_k\, h \ ,
			\end{equation*}
			where the coefficient $c_k$ is given by
			\begin{equation*}
				c_k := c_k (A, \tau ) =
                                \sum_{j=0}^{k}\, (j+1)\, \tau^j \left(
                                  \mu_2 ^{(k-1)} - \frac{1}{12}
                                \right)^{j} \ ,
			\end{equation*}
			with the second moment $\mu_2 ^{(k-1)}$
                        computed using the distribution $\rho (h)$
                        approximated at order $k-1$. Therefore one
                        simply has to solve the same equation order by
                        order, by finding at $k$-th order exactly the same solutions as in \cite{DK93} but with renormalized area 
			\begin{equation*}
				A \longmapsto \frac{2 A\, c_k }{\chi}
                                \ .
			\end{equation*}
			Since the system presents two phases \cite{DK93,Gross1994}, we have to repeat the procedure in each phase. As shown in \cite{Santilli2018}, we can go to arbitrarily high order and eventually find 
			\begin{equation*}
				c_{\infty} = \begin{cases} b_{\infty}
                                  \ , & 0<A<A_{\text{cr}} (\tau ) \ ,
                                  \\  d_{\infty} \ , & A >
                                  A_{\text{cr}} (\tau ) \ , \end{cases}
			\end{equation*}
			where the parameters $b_{\infty}$ and
                        $d_{\infty}$ are defined through the equation \cite{Santilli2018} 
			\begin{equation*}
				c_{\infty} = \left( 1- \tau\, \Big( \mu_2 - \frac{1}{12} \Big) \right)^{-2}
			\end{equation*}
			with $\mu_2$ a function of $\frac{2 A\,
                        c_{\infty}}\chi$ which is evaluated at weak
                        coupling for $b_{\infty}$ and at strong
                        coupling for $d_{\infty}$. In particular,
                        $c_{\infty} \to 1$ in the limit $\tau\to0$,
                        reproducing the undeformed case.
                        
			The critical value is found by imposing the condition \eqref{eq:condrhole1}, which is translated into the condition $\frac{ 2 A\, b_{\infty} }{\chi} < \pi^2$ on the parameters. This gives the critical line \cite{Santilli2018} 
			\begin{equation}
			\label{eq:Acr}
				A_{\text{cr}} (\tau ) =
                                \frac{\chi}{2}\, \pi ^2 \left( 1 -
                                  \tau\, \Big( \frac{1}{\pi^2} -
                                  \frac{1}{12} \Big) \right)^2 \ .
			\end{equation}
			In particular, the domain of $A$ corresponding
                        to a small coupling phase shrinks as $\tau$
                        increases, and eventually disappears at the
                        value $\tau = \tau_{\text{max}}$ defined in \eqref{eq:taumax}.
			
			Recall that in all formulas we kept track of $\chi$ for later convenience, but its physical value is $\chi=2$.

                        \subsection{Large $N$ limit of $T
                          \overline{T}$-deformed $q$-Yang-Mills theory}
                        \label{sec:TTphase1}
                        
		At this stage we are ready to apply the formalism
                reviewed in Section~\ref{sec:reviewTT1} above to
                the harder problem of $q$-Yang-Mills theory.
                
		In terms of the shifted weights $\vec h \in \mathbb{Z}^N$
                introduced in \eqref{eq:changeh}, the partition
                function of $T \overline{T}$-deformed $U(N)$ $q$-Yang-Mills theory is
		\begin{equation}
		\label{eq:ZqTTdef}
			\mz_{\textrm{\tiny $q$-YM}} ^{T
                          \overline{T}}(\lambda,\tau) = \frac{1}{N!}\,
                        \sum_{\vec h \in \mathbb{Z}^N}\, \left(  \frac{
                            \Delta_q (\vec h ) }{\Delta_{q} (  \emptyset  )
                          } \right)^{\chi } \ \exp \Bigg(  - \frac{
                          \frac{\lambda\, p }{2 N}  \, \Big(
                          \sum\limits_{j=1 } ^{N}\, h_j ^2 - \frac{
                            N\, (N^2 +1)}{12}  \Big) }{ 1 - \frac{
                            \tau}{N^3} \, \Big( \sum\limits_{j=1 }
                          ^{N}\, h_j ^2 - \frac{ N\, (N^2 +1)}{12}
                          \Big) }  \Bigg) \ ,
		\end{equation}
		where $\Delta_q(\vec h)$ is a $q$-deformation of the Vandermonde determinant
		\begin{equation*}
			\Delta_q (\vec h) = \prod_{1 \le i < j \le N}\, 2
                        \sinh \frac{ \lambda\, (h_i - h_j) }{ 2N }  \ ,
		\end{equation*}
		and we used the shorthand notation $\Delta_q (
                \emptyset ) := \Delta_q (h_i = i)$. Here $\Delta_q (
                \emptyset  )$ plays the role of a $q$-deformation of
                the Barnes $G$-function defined in
                \eqref{eq:BarnesG}. We have also continued $\chi$ beyond its physical value $\chi = \chi (\mathbb{S}^2) =2$ as in \eqref{eq:continuechi}.
		
		We now take the large $N$ limit of \eqref{eq:ZqTTdef}, which we stress is a 't~Hooft limit with 't~Hooft coupling $\lambda$, while the Yang-Mills coupling is $\lambda /N$. 
		In this limit, the contribution of $\Delta_q (
                \emptyset )^{- \chi} $ is given by~\cite{Arsiwalla}
		\begin{equation*}
			\lim_{N \to \infty }\, \chi \log  \Delta_q (
                        \emptyset ) = - \frac{\chi}{\lambda^2} \, F_0
                        ^{\textrm{\tiny CS}} (\lambda )  \ ,
		\end{equation*}
		where $F_0 ^{\textrm{\tiny CS}}(\lambda) $ is the
                planar free energy of $U(N)$ Chern-Simons theory on the
                three-sphere $\mathbb{S}^3$:
		\begin{equation*}
			F_0 ^{\textrm{\tiny CS}} (\lambda )  =
                        \frac{\lambda^3}{12}  - \frac{\pi^2}{6}\,
                        \lambda - \mathrm{Li}_3 (\e^{- \lambda } ) +
                        \zeta (3) \ .
                      \end{equation*}
                      
		The analogue of the saddle point equation \eqref{eq:SPETT} in this $q$-deformed setting is 
		\begin{equation}
		\label{eq:completeSPE}
			 \ncint\ \dd u \, \rho (u) \coth \frac{
                           \lambda \left( h-u \right) }{2} = \frac{2
                           p}{\chi}\, h\, \sum_{k=0} ^{\infty}\,
                         (k+1)\, \tau^k  \left(
                           \int_{\text{supp}(\rho)}\, \dd u\, \rho (u)\, u^2 - \frac{1}{12} \right)^k \ .
		\end{equation}
		The solution will be a function 
		\begin{equation*}
			\rho (h) := \rho \big ( h; \lambda , \tfrac{2 p}{ \chi } , \tau \big) 
		\end{equation*}
		depending parametrically on the couplings, which is
                normalized and satisfies the constraint from
                \eqref{eq:condrhole1}: $\rho (h) \le 1$. We notice
                also that $\chi $ only enters the large $N$ limit in
                the combination $\frac{2 p}{\chi}$, and hence is
                simply a rescaling of $p$.
                
		We solve the saddle point equation
                \eqref{eq:completeSPE} perturbatively, as in
                \cite{Santilli2018}  and reviewed in 
                Section~\ref{sec:reviewTT1} above. Assuming a one-cut solution, the zeroth order solution is \cite{Arsiwalla,JafferisMarsano,Caporaso2005} 
		\begin{equation*}
			\rho^{(0)} (h) = \frac{2 p}{\pi\, \chi }
                        \tan^{-1} \sqrt{ \frac{ \e^{\chi\, \lambda / 2p}
                          }{ \cosh^2 \frac{\lambda\, \chi  }{4 p }\,
                            h } -1 } \ , 
		\end{equation*}
		with support 
		\begin{equation*}
					 \mathrm{supp}
                                         \big(\rho^{(0)}\big) = \big[
                                         - \alpha^{(0)} , \alpha^{(0)}
                                         \big] \qquad \text{where}
                                         \quad \alpha^{(0)} =
                                         \frac{2}{\lambda} \cosh^{-1}
                                         \e^{\chi\, \lambda /4 p } \ .
		\end{equation*}
		We have put the superscript $^{(0)}$ everywhere to
                remind us that this is the zeroth order solution in a perturbative expansion in $\tau$. The second moment of this distribution is
		\begin{align*}
			\mu_2 ^{(0)} = \int_{- \alpha^{(0)}}
                                       ^{\alpha^{(0)}}\, \dd u\,
                  \rho^{(0)} (u)\, u^2 =  \frac{\chi^2}{12 p^2} +
                  \frac{1}{ 3 \lambda^2  } \left( \pi^2 + 6\,
                  \mathrm{Li}_2 \big( \e^{- \chi\, \lambda / 2p} \big)
                  \right) + \frac{ 8 p}{ \chi\, \lambda^3} \left(
                  \mathrm{Li}_3 \big( \e^{- \chi\, \lambda / 2p} \big)
                  - \zeta (3) \right) \ .
		\end{align*}
                
		The next order approximation of \eqref{eq:completeSPE} is 
		\begin{equation*}
		 \ncint\ \dd u \, \rho (u) \coth \frac{ \lambda \left( h-u \right) }{2} = \frac{2 p}{\chi}\, b_{q,1}\, h \ ,
		\end{equation*}
		where 
		\begin{equation*}
			b_{q,1} := b_{q,1} \big( \lambda, \tfrac{2p}{\chi} , \tau
                        \big) = 1 + \tau \left( \mu_2 ^{(0)}  -
                          \frac{1}{12} \right) \ .
		\end{equation*}
		The solution at this order then will be again as in
                \cite{Arsiwalla,JafferisMarsano,Caporaso2005}, but
                with a renormalized value of $p$ given by
		\begin{equation*}
			p \longmapsto \frac{2 p}{\chi}\, b_{q,1} \ .
		\end{equation*}
		Iterating this argument, at a generic order $O
                (\tau^{k})$ the saddle point equation is the same but the renormalization of $p$ at this order is 
		\begin{equation*}
			\frac{2 p}{\chi}\, b_{q,k} \ .
		\end{equation*}
		The parameter $b_{q,k} $ is obtained using the approximation $\mu_2 ^{(k-1)}$, which is itself a function of $b_{q,k-1}$. We find 
		\begin{equation*}
			b_{q,k} = \frac{ \dd \ }{\dd x }\left.\left(
                            \frac{1 - x^{k+1} }{1-x} - 1
                          \right)\right|_{x = \tau \, \big( \mu_2
                          ^{(k-1)} - \frac{1}{12} \big) } \ ,
		\end{equation*}
		and the convergence at $k \to \infty$ is guaranteed by
                the convergence of the geometric series defining the
                $T \overline{T}$-deformation. Although the study of
                the limiting value $b_{q,\infty}$ is based on exactly
                the same arguments as for $b_{\infty}$ in
                \cite{Santilli2018}, it is difficult to find explicit
                formulas due to the $q$-deformation. We provide more
                details on $b_{q,\infty}$ and an approximate study in
                the large $p$ regime in Appendix \ref{app:solbqinfty}.
                
		Even without an explicit expression, we can extract information about $b_{q, \infty}$ from its defining equation
		\begin{equation}
		\label{eq:finftycondition}
			b_{q, \infty} = \left( 1 - \tau\, \Big(  \mu_2
                          - \frac{1}{12} \Big) \right)^{-2} \ ,
		\end{equation}
		with $\mu_2$ depending itself on $b_{q, \infty}
                $. From this equation we already see that $b_{q,
                  \infty} \ge 1$, with equality only at $\tau=0$.
                
		We also have to check the consistency of the $T \overline{T}$-deformation. Looking back at \eqref{eq:C2toTT}, we have to find for what values of $\tau$ the inequality
		\begin{equation*}
			\tau \left( \mu _2  - \frac{1}{12} \right) < 1
		\end{equation*}
		is satisfied, so that the deformed Casimir is a
                well-defined (positive) deformation of the quadratic
                Casimir of $U(N)$. From \eqref{eq:finftycondition},
                the left-hand side of this inequality is $1 - 1/\sqrt{ b_{q, \infty}}$, with $b_{q, \infty} \ge 1$, hence the $T \overline{T}$-deformation is well-posed for all non-negative values of $\tau$. Thus the theory is well-defined at large $N$ all along the RG flow triggered by the $T \overline{T}$-deformation. This was not obvious from \eqref{eq:C2toTT} and we regard it as a strong consistency check.
		
		\subsection{Critical curves}
		\label{eq:critical}
		
		We now set $\chi$ equal to its physical value $\chi =
                \chi (\mathbb{S}^2)=2$.
                
		The solution we found in Section~\ref{sec:TTphase1}
                above holds as long as $\rho$ satisfies the
                requirement \eqref{eq:condrhole1}. From the formulas
                above, and the property $|\tan^{-1}(x)|\leq\frac\pi2$, we have 
		\begin{equation*}
			\rho (h) < \frac{ p\, b_{q, \infty} }{2}
		\end{equation*}
		and the system undergoes a phase transition only for
                those values of $p> p_0 $, where $p_0 := p_0 (\tau)$
                is implicitly defined by
		\begin{equation*}
			p_0 \, b_{q, \infty} ( p_0, \tau) = 2 \ .
		\end{equation*}
		At $\tau=0$, this was used in
                \cite{Arsiwalla,JafferisMarsano,Caporaso2005} to show
                that for $p \le \chi (\mathbb{S}^2)=2$ there is only
                one phase, while two phases separated by a critical
                line in the $(\lambda, p)$-plane appear for $p>2$.
                
		We have seen in Section \ref{sec:TTphase1} above that
                $b_{q, \infty} \ge 1$, which implies that for all
                $\tau>0$ a phase transition takes place whenever
                $p>p_0$ with $p_0<2$. As soon as the $T
                \overline{T}$-deformation is turned on, the theory
                with $p=2$ develops a strong coupling phase, with
                critical line descending from $\infty$ to a finite
                value of $\lambda$.
                
\begin{figure}[htb]
		\centering
		\includegraphics[width=0.5\textwidth]{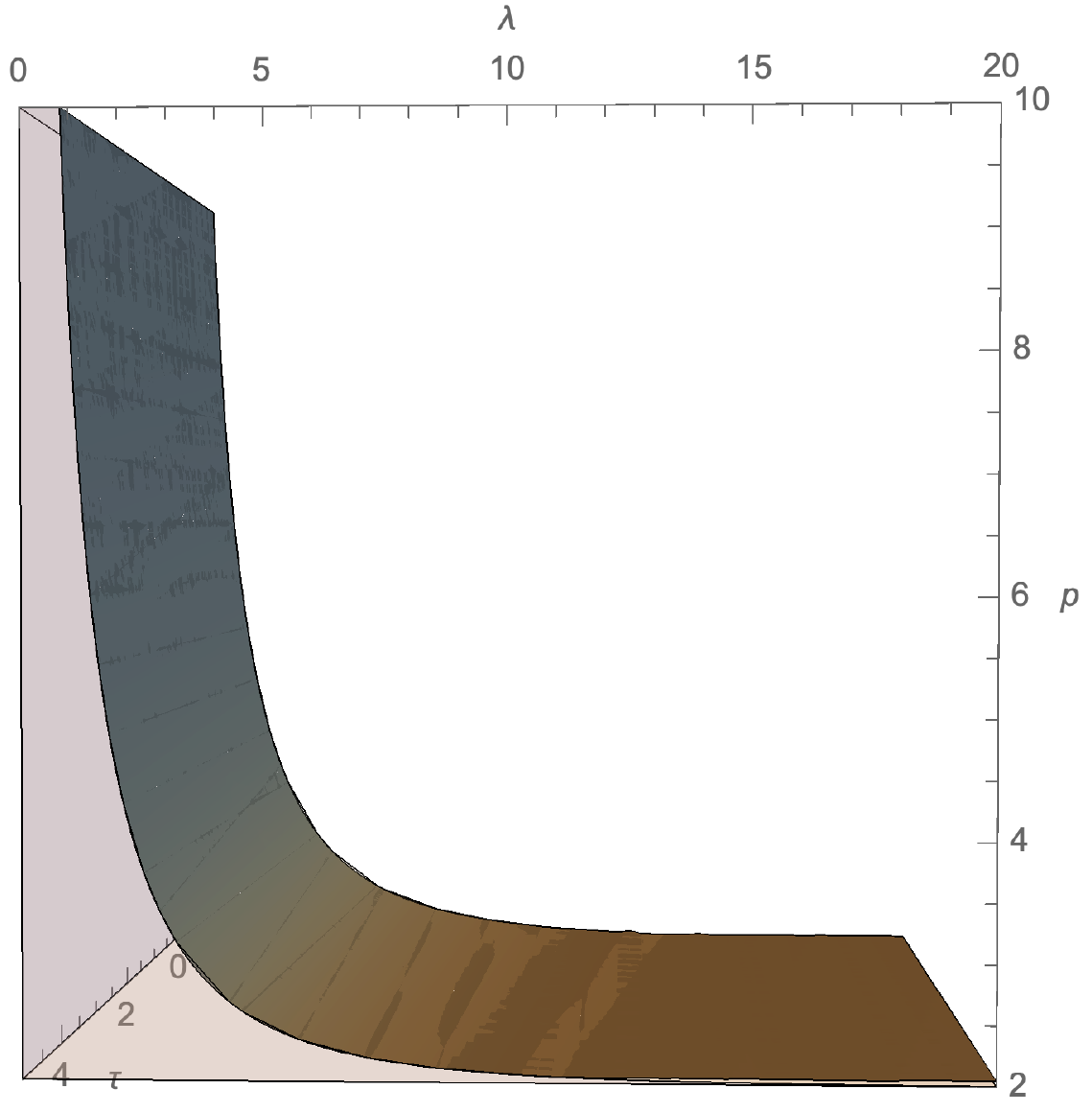}
		\caption{\small Schematic plot of the critical surface. The gray region represents the weak coupling phase.}
		\label{fig:critsurf}
		\end{figure}

		For $p> p_0 (\tau)$, the theory presents two phases,
                separated by a codimension~one critical surface in the
                octant $(\lambda >0, p>0, \tau \ge 0 )$, parametrized
                by $\lambda = \lambda_{\text{cr}} (p, \tau )$. This
                surface should be seen as a one-parameter family of
                critical curves, parameterized by $\tau \ge 0$,
                describing the evolution of the critical curve of
                Figure \ref{fig:crit-qdef} along the RG flow induced
                by the $T \overline{T}$-deformation, with $\tau$
                playing the role of the ``time''. See Figure
                \ref{fig:critsurf} for a schematic picture. (Note that
                Figure \ref{fig:critsurf} represents just a rough
                illustration of the actual critical surface.)

		This critical surface is defined implicitly by the equation 
		\begin{equation}
		\label{eq:lambdacr}
			\lambda_{\text{cr}} = p\, b_{q, \infty} \log
                        \left( 1 + \tan^2 \frac{ \pi}{p\, b_{q,
                              \infty}}  \right) \ .
		\end{equation}
		It is important to bear in mind that $b_{q, \infty}$
                depends on $\lambda$, and it must be evaluated at the
                critical value $\lambda_{\text{cr}}$ in the right-hand
                side of \eqref{eq:lambdacr}.
                
		Without an explicit expression for $b_{q, \infty}$ we
                cannot provide a formula for $\lambda_{\text{cr}} =
                \lambda_{\text{cr}} (p, \tau)$ describing the critical
                surface. Nevertheless, the lessons learned from the study of Appendix~\ref{app:solbqinfty} are that $b_{q, \infty}$ is a monotonically decreasing function of $p$, eventually approaching $b_{\infty}$ from above as $p \to \infty$. In conclusion, from Appendix \ref{app:solbqinfty} we find that $b_{q, \infty}$ decreases with $p$, and from \eqref{eq:finftycondition} we see that $b_{q, \infty}$ increases with $\tau$. This matches precisely with the known effects of the two deformations taken separately. This, together with \eqref{eq:lambdacr}, implies 
		\begin{equation*}
			p^{-1}\, A_{\text{cr}} ( \tau) \le
                        \lambda_{\text{cr}} ( p, \tau) \le
                        \lambda_{\text{cr}} ( p, 0) \ , 
		\end{equation*}
		which means that the volume in the octant $(\lambda >
                0, p >0 , \tau \ge 0)$ of the parameter space
                describing a weak coupling phase is reduced, relative
                to only the $T \overline{T}$-deformation, and is
                enhanced relative to only the $q$-deformation.
		
		\subsubsection*{$\boldsymbol{p=1}$
                  case}
                
		Since $b_{q, \infty}$ is an increasing function of
                $\tau$, one may expect that, for $\tau$ sufficiently
                large, the phase transition also takes place at
                $p=1$. In other words, one may wonder whether
                eventually $p_0 (\tau)<1$ for sufficiently large
                $\tau$. Unfortunately, our analysis of $b_{q, \infty}$
                is not reliable in the limit $p \to 1$, and we cannot
                draw any conclusions in this direction.
                
		The parameter $p$ has a geometric meaning as the
                degree of the holomorphic line bundle 
		\begin{equation*}
			\mathcal{O} (-p) \longrightarrow \Sigma \ .
		\end{equation*}
		For every $p>1$, the total space is a resolution of
                the Kleinian singularity $\mathbb{C}^2/\mathbb{Z}_p$,
                singling out $p=1$ as a special case. See also
                \cite{Szabo:2009vw} for a discussion on the physical
                relevance of $p>1$.
                
		On the other hand, there is nothing special about $p=1$ compared to $p=2$ in the original construction of \cite{Vafa2004,Aganagic2004}. It would be interesting to understand better the fate of the theory for $p=1$ with $T \overline{T}$-deformation.

		\subsection{Instanton analysis}
		\label{sec:instantons}
	
			A classic result of two-dimensional Yang-Mills theory
                        on $\cs^2$ is that the phase transition is
                        triggered by instantons \cite{Gross1994}; by
                        `instanton' here we mean a solution to the
                        classical Yang-Mills equation of motion which
                        is gauge-inequivalent to the trivial
                        connection. In
                        the weak coupling phase, the Boltzmann weight
                        of the saddle point configuration dominates
                        the partition function at large $N$. However,
                        beyond a critical value of the coupling,
                        non-perturbative contributions cease to be
                        suppressed and compete with the Boltzmann
                        weight, inducing a phase transition.
                        
			The same mechanism is at work in the $T
                        \overline{T}$-deformed
                        theory~\cite{Santilli2018}. Here the deformation reduces the suppression factor of the instantons, and therefore the phase transition takes place at lower values of the coupling.
			
			In $q$-Yang-Mills theory, again the unstable
                        instantons are the cause of the phase
                        transition~\cite{Arsiwalla,Caporaso2005}. Here
                        we show that the same arguments apply to the
                        $T \overline{T}$-deformed theory. We find
                        that, as in the situation without
                        $q$-deformation, instantons are less
                        suppressed in the $T \overline{T}$-deformed
                        theory.
                        
			We start by rewriting the sphere partition function as 
			\begin{equation}
			\label{eq:Zinstantonexp}
				\mz_{\textrm{\tiny $q$-YM}} ^{T \overline{T}}(\lambda,\tau) = \frac{1}{N!}\, \sum_{\vec\ell \in \Z^N }\, Z_{\vec\ell\,}(\lambda,\tau) \ ,
			\end{equation}
			where $Z_{\vec\ell\,}$ encodes the
                        instanton contributions, and can be obtained
                        from a modular transformation of the partition
                        function written in the representation
                        basis. See Appendix \ref{app:qYMinstantons}
                        for further details. The complete analysis for
                        $q$-Yang-Mills theory was carried out in \cite{Arsiwalla,Caporaso2005}, and we show how it is adapted to the $T \overline{T}$-deformed theory in Appendix \ref{app:qYMinstantons}. Although we cannot get a closed expression, we show that it can be evaluated order by order in $\tau$.
			
			As pointed out already in \cite{Gross1994}, focusing on the first instanton sector gives clearer insights into understanding how the non-perturbative effects kick in.
			We therefore consider the one-instanton
                        sector, for which
			\begin{equation*}
				\vec\ell = (\ell_1, 0, \dots, 0 )
                                \qquad \mbox{with} \quad \ell_1 =
                                \pm\, 1 \ .
			\end{equation*}
			Its contribution to the partition function
                        $\mz_{\textrm{\tiny $q$-YM}} ^{T \overline{T}}$ is 
			\begin{align*}
				Z_{\left( \ell_1, 0, \dots, 0
                          \right)}(\lambda,\tau) & =
                                                   \frac{1}{\Delta_q (
                                                   \emptyset )^{2}}\,
                                                   \int_{\R^N}\,
                                                   \dd\vec h\,  \e^{ -
                                                   2 \pi \,\I\,
                                                   \ell_1\, h_1  } \
                                                   \prod_{1 \le i < j
                                                   \le N} \, 4 \sinh^2 \frac{ \lambda  \left( h_i  - h_j  \right) }{2 N} \\
					& \hspace{4cm} \times \ \exp
                                   \Bigg(- \frac{\lambda\, p}{2N}\,
                                   \frac{ \sum\limits_{i=1} ^N\,  h_i
                                   ^2 - \frac{N\, (N^2 -1)}{12} }{  1
                                   - \tau\, \Big( \sum\limits_{i=1} ^N
                                   \, h_i ^2 - \frac{N\, (N^2 -1)}{12}
                                   \Big) } \Bigg) \ .
			\end{align*}
			In the large $N$ limit, the contribution of
                        $\ell_1$ to the eigenvalue density is of order
                        $O(N)$, hence sub-leading against the $O(N^2)$
                        contributions from the rest of the
                        action. Therefore in the large $N$ 't~Hooft
                        limit we can integrate over the eigenvalues
                        $h_2, \dots, h_N$ using the eigenvalue density
                        $\rho (h)$ found in Section~\ref{sec:TTphase1}
                        above. Rescaling the integration variable $h_1$ to $h= h_1 /N$, we obtain 
			\begin{equation*}
					Z_{\left( \ell_1, 0, \dots, 0
                                          \right)}(\lambda,\tau) =
                                        \frac{\mz_{N-1}(\lambda,\tau) }{\Delta_q (
                                          \emptyset )^{2}}\, \int_{\R}\,
                                        \dd h\, \e^{ -N\,
                                          S_{\text{eff}}[h] } \ ,
			\end{equation*}
			where $\mz_{N-1}$ comes from integrating out
                        the remaining $N-1$ eigenvalues, and is equal
                        to the partition function of the $U(N-1)$
                        theory in the zero-instanton
                        sector.\footnote{The symmetry breaking $U(N)
                          \to U(1) \times U(N-1)$ in the one-instanton
                          sector is explained in Appendix
                          \ref{app:qYMinstantons}.} The effective
                        action functional is given by
			\begin{equation*}
				S_{\text{eff}}[h] = - 2\,
                                \int_{\text{supp}(\rho)}\, \dd u\,
                                \rho (u) \log \left\lvert \sinh \frac{
                                    \lambda \left( h-u \right) }{2}
                                \right\rvert + \frac{\lambda\, p\,
                                  b_{q, \infty} }{2}\, h^2 - 2 \pi
                                \,\I\, \ell_1\, h \ , 
			\end{equation*}
			where we used the definition
                        \eqref{eq:finftycondition} of $b_{q, \infty}$
                        to simplify the expression. We obtain the saddle
                        point equation for the first eigenvalue given by
			\begin{equation}
			\label{eq:saddlepoint1inst}
				p\, b_{q, \infty} \, h - \frac{ 2 \pi
                                  \,\I\, \ell_1}{\lambda} = \ncint\
                                \dd u\, \rho (u) \coth \frac{ \lambda
                                  \left( h-u \right) }{2} \ .
			\end{equation}
			This is a saddle point equation for $h$, with
                        $\rho (u)$ known. At this point we notice
                        that, as expected, \eqref{eq:saddlepoint1inst}
                        is the same equation found in
                        \cite{Arsiwalla,JafferisMarsano}, except for
                        the renormalization $p \mapsto p\,
                        b_{q,\infty}$. We can therefore read off the
                        solution from \cite{Arsiwalla,JafferisMarsano}
                        to get
			\begin{equation}
			\label{eq:casesinstantonsolution}
				h = \begin{cases}  \frac{2 \,\I\,
                                    \ell_1 }{\lambda} \tan^{-1} \sqrt{
                                    \frac{ \e^{- \lambda/p\, b_{q,
                                          \infty} }  }{ \cos^2 \frac{
                                        \pi }{p\, b_{q, \infty} } } -1
                                  } \ , &  p\, b_{q, \infty} > 2 \ ,
                                  \\ \frac{ 2 \pi \,\I\, \ell_1
                                  }{\lambda} \ , &  p\, b_{q, \infty}
                                  \le 2 \ , \end{cases}
			\end{equation}
			where we used $\lvert \ell_1 \rvert = 1$ to simplify
			\begin{equation*}
				\cosh \left( \frac{ \lambda}{2 p\,
                                    b_{q, \infty}}\, \frac{ 2 \pi
                                    \,\I\, \lvert \ell_1 \rvert
                                  }{\lambda} \right) = \cos
                                \frac{\pi}{p\, b_{q, \infty}} \ .
                              \end{equation*}
                              
			From \eqref{eq:casesinstantonsolution} it follows that there is no phase transition for $p$ below a critical value $p_0 (\tau)$, defined such that
			\begin{equation*}
				p\, b_{q,\infty} (\lambda, p, \tau )
                                \le 2 \qquad \mbox{for} \quad \lambda
                                >0 \qquad \text{when} \quad p \le p_0
                                (\tau) \ ,
			\end{equation*}
			because then the instanton contributions are
                        suppressed for all values of $\lambda$. On the
                        other hand, when $p>p_0(\tau) $ and $p\, b_{q,
                          \infty} >2$ we have (dropping an irrelevant
                        overall constant)
			\begin{equation*}
			\frac{ Z_{\left( \ell_1, 0, \dots, 0 \right)}(\lambda,\tau) }{ Z_{\left( 0, 0, \dots, 0 \right)}(\lambda,\tau) } = \exp \left( - \frac{N}{\lambda\, p }\, \gamma( \lambda, p\, b_{q, \infty}) \right)
			\end{equation*}
			where $\gamma$ is the function defined in \cite{Arsiwalla}, which in turn is a one-parameter deformation of the suppression function found in \cite{Gross1994}. When 
			\begin{equation*}
				\frac{ \e^{ - \lambda/p\, b_{q,
                                      \infty} }  }{ \cos^2 \frac{ \pi
                                  }{p\, b_{q, \infty} } } - 1 >0 \ ,
			\end{equation*}
			corresponding to $\lambda <
                        \lambda_{\text{cr}}$, the function $\gamma$ is
                        a positive decreasing function of $\lambda$,
                        for any fixed $p$. However, it becomes purely
                        imaginary when $\lambda >
                        \lambda_{\text{cr}}$, implying that the
                        one-instanton sector is no longer suppressed
                        and its contribution becomes relevant. 
		
		\subsection{Strong coupling phase}
		\label{sec:TTphase2}
		
		For values of $\lambda$ such that, for given $p$ and
                $\tau$, the eigenvalue density $\rho (h)$ found in
                Section~\ref{sec:TTphase1} above does not satisfy the
                constraint \eqref{eq:condrhole1}, we have to drop the
                assumption of a one-cut solution and find another
                distribution $\rho (h)$ satisfying the bound.
                
		The strategy is the same as that followed in
                Section~\ref{sec:TTphase1} above for the weak coupling
                phase: expanding the saddle point equation
                \eqref{eq:completeSPE} as a power series in $\tau$, we
                can solve it iteratively. We do not spell out the
                technical details here, as they are exactly as in
                \cite{Arsiwalla,JafferisMarsano,Caporaso2005}, up to
                the renormalization $p \mapsto \frac{2 p}{\chi}\,
                d_{q,k}$. The coefficient 
		\begin{equation}
		\label{eq:defdqk}
			d_{q, k} = \sum_{j=0} ^k\, (j+1)\, \tau^{j}
                        \left( \mu_2 ^{(k-1)} - \frac{1}{12}
                        \right)^{j}
		\end{equation}
		is formally the same as $b_{q,k}$, but they differ in
                that $b_{q,k}$ is computed using the second moment
                $\mu_2$ of the distribution $\rho (h)$ at weak
                coupling in \eqref{eq:finftycondition}, whereas for
                $d_{q,k}$ the moment $\mu_2$ corresponds to $\rho (h)$
                at strong coupling in \eqref{eq:defdqk}. We
                distinguish the weak and strong coupling solutions by
                $\rho_{\rm weak}$ and $\rho_{\rm strong}$.
		The complete solution would require finding $d_{q,
                  \infty}$. 
				
		\subsubsection*{Third order phase transition}
		
		Even without closed expressions available for $b_{q, \infty}$
                and $d_{q, \infty}$, we can extract information from
                the form of the eigenvalue density and from what is known
                at $\tau=0$. In fact, at $\tau=0$ the phase transition
                is of third order
                \cite{Arsiwalla,JafferisMarsano,Caporaso2005}, which
                means that $\log \mathcal{Z}_{\textrm{\tiny $q$-YM}} $
                is twice continuously differentiable along the
                critical curve $\lambda = \lambda_{\text{cr}} (p)$.
                
		The crucial feature is that one can extract the
                derivative $\frac{\partial \ }{\partial \lambda } \log
                \mathcal{Z}_{\textrm{\tiny $q$-YM}} ^{T \overline{T}}
                $ from the second moment $\mu_2 [\rho]$ in each phase,
                and a third order phase transition implies that $\mu_2
                [\rho_{\text{weak}}]$ and $\mu_2
                [\rho_{\text{strong}}]$ agree up to their first
                derivatives at $\tau=0$. We now exploit this fact to
                describe the behaviours of $b_{q, \infty}$ and $d_{q,
                  \infty}$ close to the critical curve. Using the
                defining expressions \eqref{eq:finftycondition} and
                \eqref{eq:defdqk} for $b_{q, \infty}$ and $d_{q,
                  \infty}$, we can expand in $\lambda$ around
                $\lambda= \lambda_{\text{cr}}$ to get
		\begin{align*}
			b_{q, \infty} & = f_{0} ^{\text{weak}} (b_{q,
                                        \infty}
                                        \vert_{\lambda=\lambda_{\text{cr}}
                                        } ) + \left( \lambda -
                                        \lambda_{\text{cr}}
                                        \right)^2\, f_{2}
                                        ^{\text{weak}} (b_{q, \infty}
                                        \vert_{\lambda=\lambda_{\text{cr}}
                                        } ) + O \big((\lambda -
                                        \lambda_{\text{cr}})^3\big) \ , \\[4pt]
			d_{q, \infty} & = f_{0} ^{\text{strong}}
                                        (d_{q, \infty}
                                        \vert_{\lambda=\lambda_{\text{cr}}
                                        } ) + \left( \lambda -
                                        \lambda_{\text{cr}}
                                        \right)^2\, f_{2}
                                        ^{\text{strong}}  (d_{q,
                                        \infty}
                                        \vert_{\lambda=\lambda_{\text{cr}}
                                        } ) + O \big((\lambda -
                                        \lambda_{\text{cr}})^3\big) \ .
		\end{align*}
		The agreement of $\mu_2$ at weak and strong coupling
                at $\tau=0$ can be used to show that $f_{0}
                ^{\text{weak}} = f_{0} ^{\text{strong}} $ at the
                critical point for all $\tau \ge 0$, which in turn
                guarantees that $b_{q, \infty}$ and $d_{q, \infty}$
                agree up to the first derivative. Essentially, by
                direct inspection one finds that the defining equation
                for $d_{q, \infty}$ is exactly the same as for $b_{q,
                  \infty}$ at order $O(\lambda -
                \lambda_{\text{cr}})$.
                
		On the other hand, one can check that, after inclusion
                of the renormalization of $p$ at weak or strong
                coupling, the first derivative $\frac{\partial \
                }{\partial \lambda } \log \mathcal{Z}_{\textrm{\tiny
                    $q$-YM}} ^{T \overline{T}} $ depends only on
                $b_{q, \infty}$ (respectively $d_{q, \infty}$) at weak
                (respectively strong) coupling, and not on their
                derivatives. This is done again by expanding the
                integral formula for $\mu_2$ close to
                $\lambda_{\text{cr}}$, and is a consequence of the
                very simple way in which the parameters $b_{q,\infty}$
                and $d_{q,\infty}$ enter.
                
		As an immediate consequence, the second derivative
                $\frac{\partial^2 \ }{\partial \lambda^2 } \log
                \mathcal{Z}_{\textrm{\tiny $q$-YM}}^{T \overline{T}}
                \big\vert_{\lambda=\lambda_{\text{cr}}}$ depends only
                on the 1-jets of $b_{q, \infty}$ and $d_{q,
                  \infty}$ at $\lambda_{\text{cr}}$, which we have
                argued to match. We therefore find that the phase
                transition is of third order.
                
		We refer to \cite{Santilli2018} for further discussion, since the details of the argument do not depend on the explicit form of $\rho (h)$ once we zoom in close to the critical curve.

		\subsection{Refinement}
		\label{sec:tqTTLargeN}
		
		Consider the refinement of $U(N)$ $q$-Yang-Mills
                theory \cite{Aganagic2012} with refinement parameter
                $t=q^{\beta}$, for $\beta \in \mathbb{Z}_{> 0}$; the
                unrefined limit $t=q$ then corresponds to $\beta=1$. A
                refined definition of the shifted weight variables
                $\vec h
                \in \mathbb{Z}^{N}$ introduced in \eqref{eq:changeh}
                is given by
		\begin{equation*}
			h_i = - R_i + \beta \left( i - \frac{N+1}{2}
                        \right) \qquad \mbox{for} \quad i =1, \dots, N
                        \ .
		\end{equation*}
		 In this basis we find that the $T \overline{T}$-deformed partition function on $\mathbb{S}^2$ is 
		 \begin{equation}
		 \label{eq:ZqtYMTT}
		 	\mz_{\textrm{\tiny $(q,t)$-YM}} ^{T
                          \overline{T}}(\lambda,\tau) = \frac{1}{N!}\,
                        \sum_{\vec h \in \mathbb{Z}^N }\, \frac{
                          \Delta_{(q,t)} (\vec h) \, \Delta_{(q,t)}
                          (-\vec h) }{ -\Delta_{(q,t)} ( \emptyset )^2
                        } \ \exp \Bigg( - \frac{ \frac{\lambda\,
                            p}{2N} \Big( \sum\limits_{i=1}^{N}\,
                          h_{i} ^2 - \beta^2\, \frac{N\, (N^2-1)}{12}
                          \Big) }{ 1 - \frac{\tau}{N^3}\,  \Big(
                          \sum\limits_{i=1}^{N} \, h_{i} ^2 -
                          \beta^2\, \frac{N\, (N^2-1)}{12}  \Big)  }
                        \Bigg) \  ,
		 \end{equation}
		 in which the $T \overline{T}$-deformation only
                 changes the refined (or $\beta$-deformed) quadratic
                 Casimir, relative to the case $\tau=0$. 
		 The Macdonald measure $\Delta_{(q,t)} (\vec h) $ in \eqref{eq:ZqtYMTT} is given for $t= q^{\beta}$ as 
		 \begin{equation*}
		 	\Delta_{(q,t)} (\vec h) = \prod_{m=0}
                        ^{\beta-1} \ \prod_{1 \le i < j \le N}\,  2
                        \sinh \frac{ \lambda \left(  h_i - h_j  + m
                          \right) }{ 2N } \ ,
		 \end{equation*}
		 and $\Delta_{(q,t)} ( \emptyset ) $ is obtained by
                 setting $h_i - h_j = \beta\, (i-j)$. This discrete
                 matrix model has been thoroughly studied in~\cite{Szabo2013}.
		
		It was proven in \cite{KoSiSz} that the 't Hooft limit
                of $(q,t)$-Yang-Mills theory coincides with that of
                $q$-Yang-Mills theory with rescaled coupling
                $\lambda^{\prime} = \beta\, \lambda $. The proof
                of \cite{KoSiSz} is straightforwardly adapted to the
                $T \overline{T}$-deformed setting, and we find that
                $q$-Yang-Mills theory and its refinement coincide in the 't Hooft limit after rescaling 
		\begin{equation*}
			\lambda^{\prime} = \beta\, \lambda \qquad
                        \mbox{and} \qquad \tau^{\prime} = \beta^2\,
                        \tau \ .
		\end{equation*}
		When $\beta >1$ the refinement non-trivially modifies the deformation parameter $\tau$, and the weak coupling region in parameter space is drastically reduced relative to the unrefined case $\beta=1$.
	
	\section{Entanglement entropy}
	\label{sec:entanglement}	

		The goal of this section is to study the von~Neumann entanglement
                entropy of $T \overline{T}$-deformed Yang-Mills
                theory~\cite{DonnellyShyam1} in a
                general state at large
                $N$. The entanglement entropy of two-dimensional
                Yang-Mills theory has been studied
                in~\cite{GromovSantos,Donnelly2014} at finite $N$ (see
                also~\cite{Donnelly2016,Donnelly2018}), and a thorough
                analysis of the large $N$ limit in both phases and
                including sub-leading corrections appeared
                in~\cite{Donnelly2019}. The
                $T\overline{T}$-deformation at finite $N$ was
                considered in~\cite{Ireland}. In this section we give a
                first step towards the extension of the large $N$
                analysis to the $T \overline{T}$-deformed setting.
                
		The formalism of
                \cite{Donnelly2018} used to study the entanglement entropy
                (at finite $N$) is well-suited to the class of almost
                topological gauge theories considered in the present
                paper, and can be adapted to $q$-Yang-Mills theory and
                its $T \overline{T}$-deformation. In the following we
                consider only the case of ordinary Yang-Mills theory.
		
		\subsubsection*{Entanglement entropy at large
                  $\boldsymbol N$}
                
		The calculations which follow mimic those
                of~\cite{Donnelly2019}. We start by using the replica
                trick \cite{Parisi} to write down the formula for the entanglement
                entropy. The replica trick consists in defining an
                auxiliary manifold $\Sigma^n$, which is an $n$-sheeted
                Riemann surface obtained by cyclically gluing $n$ sheets
                along cuts on a spatial subregion of $\Sigma$. This defines a branched cover
                $\Sigma^n\longrightarrow\Sigma$. One then continues the dependence
                on $n$ of the partition function of a quantum
                field theory $\mt$ on $\Sigma^n$ to values
                $n\in\R_{>0}$. We can then compute the entanglement entropy as 
		\begin{equation*}
			S_{\text{entang}}[\Sigma] =
                        \left. -\frac{\partial \ }{\partial
                            n}\bigg(\frac{\mathcal{Z}_{\mt}[\Sigma^n]}{\mathcal{Z}_{\mt}[\Sigma]^n}
                            \bigg)\right|_{n=1} = \left. \left( 1 - 
                            \frac{\partial \ }{\partial n} \right) \log
                          \mathcal{Z}_{\mt}
                          [\Sigma^n ] \, \right\rvert_{n=1} \ .
                      \end{equation*}
    This formula may seem problematic, due to the non-uniqueness of analytic continuation from $n \in \mathbb{Z}_{>0}$ to $n \in \R_{>0}$. Usually, one would need to define a continuation to $n \in \mathbb{C}$ and then check, with the aid of Carlson's theorem, that different continuations would differ by an everywhere vanishing function. 
    Nevertheless, as will be manifest below, we can translate this
    problem into the problem of analytically continuing the Euler
    characteristic $\chi (\Sigma)$ to a real quantity $\chi$ in a
    neighbourhood of the physical value $\chi=2$. Throughout Section
    \ref{sec:phaseTTqYM} we have been careful in keeping track of the
    Euler characteristic $\chi$ in all the formulas, and we can
    continue them to $\chi >0$.\footnote{We stress that we are
      computing the entanglement entropy of a quantum field theory, in which JT gravity has been integrated out \emph{ab initio}. Therefore the replicas that appear here should not be confused with the (one-dimensional unsewn) replicas used in purely gravitational theories.}
                      
		Following \cite{Donnelly2019}, in the case at hand we define 
		\begin{equation*}
			P (\vec h) =
                        \frac{1}{\mathcal{Z}_{\textrm{\tiny YM}} ^{T
                            \overline{T}}(A,\tau)} \, \bigg(
                          \frac{\Delta (\vec h) }{\Delta (\emptyset )}
                        \bigg)^{\chi}\, \e^{- \frac{A}{2N}\, C_2 ^{T
                            \overline{T} } (\vec h,\tau)}
		\end{equation*}
		which is the probability of finding a given configuration of weights
                $\vec h \in \mathbb{Z}^{N}$, with variables as defined
                in \eqref{eq:changeh}, where we recall that $\Delta
                (\vec h)$ is the Vandermonde determinant \eqref{eq:Vandermonde}. Note that $\Delta (\emptyset
                )^{\chi} = G(N+1)^{\chi}$ does not contribute, since
                it is cancelled by the normalization.  We see that there is no issue in defining $P(h)$ for non-integer $\chi >0$.\par
        Then the replica trick gives~\cite{Donnelly2019} 
		\begin{equation*}
			S_{\text{entang}}(A,\tau) = -
                          \sum_{\vec h \in \mathbb{Z}^N}\, P (\vec h) \log
                          P(\vec h) + \sum_{\vec h \in \mathbb{Z}^N}\,
                          P (\vec h) \log \Delta (\vec h) \ ,
		\end{equation*}
		where the first series is identified with the Shannon
                entropy, which measures the entropy due to the
                fluctuations of the values of the weights $\vec h$, while the second series is identified with the Boltzmann
                entropy, which measures the total entropy due to each
                sector $\vec h$ weighted with their probability (see~\cite{Donnelly2014,Donnelly2016,Donnelly2019} for
                further details).

                We notice that the Shannon entropy is
		\begin{equation*}
			S_{\text{Shan}}(A,\tau) := - \sum_{\vec h \in \mathbb{Z}^N}\, P (\vec h) \log
                          P(\vec h) = \sum_{\vec h \in \mathbb{Z}^N}\,
                          P(\vec h)\,
                          S_{\textrm{\tiny YM}} [\vec h] \ .
		\end{equation*}
		This implies in particular that at large $N$, when the saddle point configuration dominates and the fluctuations are averaged out, we get 
		\begin{equation*}
			\lim_{N\to\infty}\, S_{\text{Shan}} (A, \tau)
                        = 0 \ .
		\end{equation*}
		This was also found in \cite{Donnelly2019}, and it is
                in fact a general feature of all the deformations of
                two-dimensional Yang-Mills theory studied in this
                paper, which follows directly from their matrix model
                presentations. This of course will not hold after the
                introduction of sub-leading corrections.

                In the remainder of this section we focus on the
                Boltzmann entropy
                \begin{align*}
S_{\text{Boltz}}(A,\tau) := \sum_{\vec h \in \mathbb{Z}^N}\,
                          P (\vec h) \log \Delta (\vec h) \ .
                \end{align*}
                
		\subsubsection*{Entanglement entropy in the weak
                  coupling phase}
                
			Let us define 
			\begin{equation*}
				\mathcal{F}^{T \overline{T}}(A,\tau)
                                := \lim_{N \to \infty }\, \frac{1}{N^2}
                                \log \mathcal{Z}_{\textrm{\tiny YM}}
                                ^{T \overline{T}}(A,\tau) \  .
			\end{equation*}
			This quantity is relevant for the computation
                        of the entanglement entropy at large $N$,
                        because the Boltzmann entropy normalized by
                        $N^2$ and by the number of entangling points
                        is given by
			\begin{equation}
			\label{eq:SBoltznorm}
				 \widetilde{S}_{\text{Boltz}}(A,\tau) =
                                 \frac{ \partial \mathcal{F}^{T
                                       \overline{T}}(A,\tau)  }{\partial
                                     \chi} \ .
			\end{equation}
			In the large $N$ limit we find 
			\begin{equation}\label{eq:FTTbarlargeN}
				\mathcal{F}^{T \overline{T}}(A,\tau) =
                                \frac{3 \chi}{4}  + \frac{\chi}{2} \,
                                \int_{\text{supp}(\rho)}\, \dd h\,
                                \rho (h) \log \vert h \vert -
                                \frac{A}{4} \left( \frac{ C_2 -
                                    \frac{1}{12} }{ 1 - \tau\, C_2  }
                                \right) \ ,
			\end{equation}
			where 
			\begin{equation*}
				C_2 = \mu_2 - \frac{1}{12} =
                                \int_{\text{supp}(\rho)}\, \dd h\, \rho (h)\, h^2 - \frac{1}{12} 
			\end{equation*}
			is the limiting value of the quadratic Casimir
                        at large $N$. (Note that $-\frac{1}{12}$
                        appears twice in the last numerator of \eqref{eq:FTTbarlargeN}.) This
                        expression is now to be evaluated in each phase
                        using the corresponding eigenvalue density
                        $\rho (h)$ obtained in \cite{Santilli2018}.
                        
\begin{figure}[htb]
	\centering
	\includegraphics[width=0.65\textwidth]{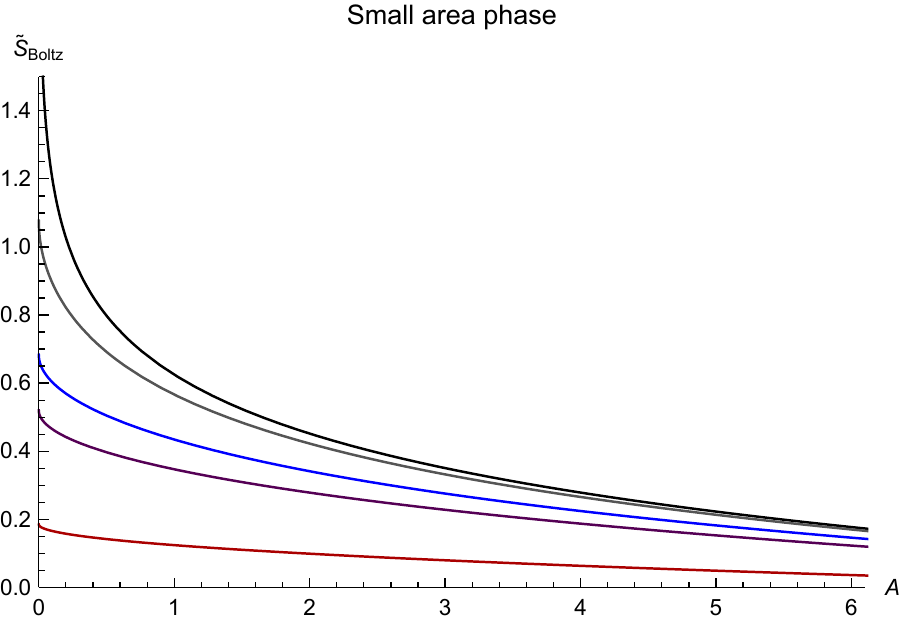}
	\caption{\small Entanglement entropy of $T
          \overline{T}$-deformed Yang-Mills theory at large $N$,
          normalized by $N^2$ and per number of entangling points, as
          a function of the area $A$ in the weak coupling phase. The
          different curves correspond to the values $\tau=0$ (black), $\tau=0.1$ (gray), $\tau=0.5$ (blue), $\tau=1$
          (purple), and $\tau=5$~(red).}
	\label{fig:BoltzTTYMweak}
	\end{figure}

			In the weak coupling (small area) phase we obtain 
			\begin{equation*}
				\mathcal{F}^{T \overline{T}}  (A <
                                A_{\text{cr}},\tau) = \frac{\chi}{2} -
                                \frac{ \chi }{4} \log \left( \frac{ 2
                                    A\, b_{\infty} }{\chi } \right) -
                                \frac{A  }{ 4  } \, \frac{
                                  \frac{\chi}{2 A\, b_{\infty}}  -
                                  \frac{1}{6} }{ 1 -  \frac{ \tau \,
                                    \chi}{2 A\, b_{\infty}} +
                                  \frac{\tau}{12} } \ ,
			\end{equation*}
			where \cite{Santilli2018} 
			\begin{equation*}
				b_{\infty} \big( \tfrac{2A}{\chi} ,
                                \tau \big) = \frac{ 1 + \frac{ \tau\,
                                    \chi }{ A}\, ( 1 + \frac{ \tau
                                  }{12} )  + \sqrt{ 1 + \frac{ 2
                                      \tau\, \chi}{A}\, (1 +
                                    \frac{\tau}{12} ) } }{ 2 (1 +
                                  \frac{ \tau}{12} )^2 } \ .
			\end{equation*}
			Plugging this expression into \eqref{eq:SBoltznorm} we obtain 
			\begin{align*}
			 \widetilde{S}_{\mathrm{Boltz}} ( A <
                          A_{\text{cr}},\tau)  & = -\frac{ A}{4}
                                                 \left( 1 -
                                                 \frac{\tau}{12}
                                                 \right) \Bigg( \frac{
                                                 \big(2 A\,
                                                 b_{\infty}(\frac{2A}\chi,\tau)
                                                 \big)^{-1} + \chi^{-1}\,
                                                 b_{\infty}(\frac{2A}\chi,\tau)
                                                 ^{-2}\, \frac{
                                                 \partial b_{\infty}
                                                 (A, \tau) }{ \partial
                                                 A } }{ \left( 1 -
                                                 \tau \big(
                                                 \frac{\chi}{2A\,
                                                 b_{\infty}(\frac{2A}\chi,\tau) }
                                                 - \frac{1}{12} \big)
                                                 \right)^2 } \Bigg) \\
				& \quad \, + \frac{1}{2} - \frac{1}{4}
                           \log \left( \frac{2A\,
                           b_{\infty}(\tfrac{2A}\chi,\tau)}\chi \right) +
                           \frac{ 3 A\, b_{\infty}(\frac{2A}\chi,\tau)}{2} +
                           \frac{A^2}{\chi}\, \frac{ \partial b_{\infty}
                           (A, \tau) }{ \partial A} \ .
		\end{align*}
		We have not inserted the explicit expressions for
                $b_\infty$ and $\frac{ \partial b_{\infty}}{\partial
                  A}$ in order to avoid clutter, but the formula can be
                straightforwardly evaluated. The normalized Boltzmann
                entropy is plotted in Figure~\ref{fig:BoltzTTYMweak}
                for $\chi=2$ and for various values of $\tau$.
                
	This describes the evolution of the result of
        \cite{Donnelly2019} along the RG flow triggered by the $T
        \overline{T}$-deformation of two-dimensional Yang-Mills
        theory, which shows that the entropy decreases with increasing
        deformation parameter $\tau$. 
			
	\section{Outlook}
	\label{sec:out}
        
	There are many intriguing open problems that could be tackled,
        but a treatment of them is out of the scope of the present
        paper. In this final section we point out four directions for
        which the present study has laid the groundwork for further
        investigation.
	
	\subsubsection*{Pullback of $\boldsymbol{T \overline{T}}$-deformation
          to $\boldsymbol{\mathcal{N}=4}$ Yang-Mills theory}
	\label{sec:stringinterpret}
	
	From the construction of \cite{Vafa2004,Aganagic2004} (see
        also \cite{Caporaso2005,Caporaso2005chiral,Griguolo:2006kp}),
        $q$-deformed Yang-Mills theory on $\Sigma$ descends from
        $\mathcal{N}=4$ supersymmetric Yang-Mills theory living on a
        non-compact four-manifold $X_4$ which is the total space of
        the degree~$p$ holomorphic line bundle 
	\begin{equation*}
		\mathcal{O} (-p) \longrightarrow \Sigma \ .
	\end{equation*}
	To implement the boundary conditions on each fibre
        $\mathbb{C}_{z}$ over points $z \in \Sigma$, one introduces a
        group element $g (z, \bar{z}) \in G$ encoding the holonomy of
        the gauge field at infinity in $\mathbb{C}_z$:
	\begin{equation*}
		g (z, \bar{z})= \e^{ \I\, \phi (z, \bar{z}) } := \exp \bigg( \oint _{S^1 _{\infty} (z) }\,  A^{(4)} \bigg) \ ,
	\end{equation*}
	where $A^{(4)}$ is the four-dimensional gauge connection and
        $S^1 _{\infty} (z)$ is the circle at infinity in the fibre
        over $z \in \Sigma$.
        
For $p>0$ the line bundle $\mathcal{O}(-p)$ is non-trivial and
        presents $p$ topological obstructions to defining the section
        $\phi (z, \bar{z})$ globally. This produces exactly $p$ copies
        of the action
	\begin{equation*}
		\int_{\Sigma}\, \tr(\phi^2)\, \omega \ ,
	\end{equation*}
	which, together with the compactness of $\phi (z, \bar{z})$,
        deforms the topological Yang-Mills theory on $\Sigma$,
        descending in the bulk from the four-dimensional gauge theory
        on $X_4$, to $q$-deformed Yang-Mills theory \cite{Vafa2004,Aganagic2004}.\par
	From our point of view it is then natural to ask what would be
        the effect and the geometrical meaning of the $T
        \overline{T}$-deformation, once the theory is embedded into
        the four-dimensional setting. In particular, it would be
        interesting to determine whether a four-dimensional picture
        emerges when the gauge theory on $\Sigma$ is dynamically coupled to JT gravity.
	
	\subsubsection*{Two-dimensional Yang-Mills-Higgs theory}
	
In~\cite{GerShat06,GerShat07} a two-dimensional Yang-Mills-Higgs
theory was considered, and it was shown that the wavefunctions of the
$U(N)$ theory coincide with the wavefunctions of the $N$-particle
Hamiltonian of nonlinear Schr\"odinger theory. We shall now briefly
present the setup of \cite{GerShat06}.

Consider the Yang-Mills action and take its minimal supersymmetric
extension, with BRST multiplet $(A, \phi, \psi)$, which adds a term
$\int_\Sigma\, \mathrm{Tr} \left( \psi \wedge \psi \right)$ to the action. Denote by $\mathsf{Q}$
the supercharge under which the extended Yang-Mills action is
invariant. The Yang-Mills-Higgs theory is built by including the Higgs
field $\Phi$, a bosonic $\mathfrak{g}$-valued one-form in the adjoint
representation, and its fermionic superpartner $\Psi$, as well as two
pairs of scalar auxiliary fields $(\varphi_+,
\chi_{+})$ and $(\varphi_{-}, \chi_{-})$, with $\varphi_{\pm}$
Grassmann-even and $\chi_{\pm}$ Grassmann-odd $\mathfrak{g}$-valued
scalars in the adjoint representation. 
	The action is then further modified as 
	\begin{equation*}
		S_{\textrm{\tiny YM}} + \left\{ \mathsf{Q},
                  S_{\textrm{\tiny H}} \right\} \ ,
	\end{equation*}
	where \cite{GerShat06,GerShat07}
	\begin{equation*}
		 S_{\textrm{\tiny H}} = \int_{\Sigma}\, \mathrm{Tr}\left(
                   \frac{1}{2}\, \Phi \wedge \Psi + \alpha \left(
                     \chi_{+} \, \varphi_{-} + \chi_{-} \, \varphi_{+}
                   \right)\, \omega\right) \ ,
	\end{equation*}
	for arbitrary $\alpha \in \mathbb{R}$. Acting with
        $\mathsf{Q}$ one finds \cite{GerShat06} that the action is
        quadratic in all of the fields $(\psi, \Phi, \Psi, \varphi_{\pm},
        \chi_{\pm})$, so they can be integrated out and produce
        one-loop determinants. The abelianzation technique works in
        that case, and the one-loop determinants coming from the
        integration of the Higgs sector modify the partition function
        of the pure Yang-Mills theory on $\Sigma$ expressed in the
        representation basis.
        
	It would be very interesting to study the $T
        \overline{T}$-deformation of such theories: besides the
        deformation of the quadratic potential in the scalar $\phi$,
        also the quadratic term involving the fields $\varphi_{\pm}$
        and $\chi_{\pm}$ is affected by the $T
        \overline{T}$-deformation as in \eqref{eq:Vttdef}, and the
        fields can no longer be integrated out. Nonetheless, the gauge
        connection $A$ does not appear in any of the deformation
        terms, and the abelianization should work for the path
        integral over $\phi$. Yet, we are left with the path integral
        over $\varphi_{\pm}$ and $\chi_{\pm}$.
        
	On the other hand, it is well-known that $T \overline{T}$-deformation preserves integrability. Therefore, since the wavefunctions of the Yang-Mills-Higgs theory are those of an integrable quantum mechanics, it would be appropriate to ask whether the wavefunctions of the $T \overline{T}$-deformation of Yang-Mills-Higgs theory are still related to an integrable system.

	\subsubsection*{Defects}
        
Another aspect worth pursuing is the inclusion of defects. Here we
show how to $T \overline{T}$-deform two-dimensional Yang-Mills theory
with defects \cite{Muller2019}.

Let $P \longrightarrow \Sigma$ be a principal $G$-bundle and $D
\longrightarrow \Sigma $ an $\mathsf{Out} (G)$-bundle on $\Sigma$,
where $\mathsf{Out} (G)$ is the group of outer automorphisms of the
Lie group $G$. Following \cite{Muller2019} we define the $D$-twisted $G$-bundle 
	\begin{equation*}
		 P^{\prime}  \longrightarrow \Sigma \ ,  
	\end{equation*}
which is a $G^{\prime}$-bundle with structure group 
	\begin{equation*}
G^{\prime} \cong G \rtimes \mathsf{Out} (G) \ .
	\end{equation*}
The partition function  of two-dimensional Yang-Mills theory with symmetry
defects is then \cite{Muller2019}
	\begin{equation*}
		\mathcal{Z}_{\textrm{\tiny YM-def}}[\Sigma] =
                \int_{\mathscr{A}_G ^{D} (\Sigma)}\, \mathscr{D}
                (P^{\prime},A) \ \int\, \mathscr{D} \phi\, \e^{-
                  S_{\textrm{\tiny YM}} (P',A, \phi)} \ , 
	\end{equation*}
where $\mathscr{A}_G ^{D} (\Sigma)$ is the space of pairs
$(P^{\prime},A)$ of $D$-twisted $G$-bundles $P^{\prime}  \longrightarrow
\Sigma $ with connection $A$. The scalar field $\phi$ can only be
introduced after fixing $P^{\prime}$, while the action is the standard
Yang-Mills action, where now we have stressed the dependence on the
choice of $D$-twisted bundle.

As extensively discussed in Section \ref{sec:almosttop}, to turn on
the $T \overline{T}$-deformation, we must insert JT gravity as the
innermost path integral here. It is immediately seen that we can again
integrate out the dynamical coframe field, reproducing the deformation of
the potential \eqref{eq:Vttdef}. We stress that one should be very
careful with the order of path integration: since $\phi$ is only
introduced after fixing a choice of bundle $P^{\prime}$, one must
first compute the gravitational path integral, followed by the
path integral over $\phi$ as a function of the choice $(P^{\prime},A)
\in \mathscr{A}_G ^{D} (\Sigma)$, and only then eventually integrate
over $\mathscr{A}_G ^{D} (\Sigma)$. 
	
The outcome of this analysis is that $T \overline{T}$-deforming
two-dimensional Yang-Mills theory with defects modifies the action
exactly as without defects. On the other hand, it would be interesting
to investigate further the interplay between $T
\overline{T}$-deformation and symmetry defects, which requires the introduction
of defects in the already $T \overline{T}$-deformed field theory. This
would entail a reformulation of the $T \overline{T}$-deformed
quantum field theory in the formalism of \cite{Runkel2018}. 
	
\subsubsection*{$\boldsymbol{T \overline{T}}$-deformation in higher dimensions}

We conclude on a rather speculative note. Recalling that a univocal
definition of what the analogues of the $T \overline{T}$-deformation
would be in higher dimensions is still lacking, a relevant question is
the following: If we take a two-dimensional theory \smash{$\mathscr{T}_A
^{(2)}$} with a known relationship to a higher-dimensional field theory
\smash{$\mathscr{T}_B ^{(d>2)}$} and deform it, what would be the effect on
\smash{$\mathscr{T}_B ^{(d>2)}$}? This question can be represented
symbolically by the following diagram:
	\begin{equation}
		\begin{tikzpicture}[auto,node distance=1.5cm,baseline=-1.4cm]
				\node[] (TA) {$ \mathscr{T}_A ^{(2)} \ $};
				\node[] (TB) [right = of TA]{$\mathscr{T}_B ^{(d>2)} $};
				\node[] (TTTA) [below = of TA]{$ \mathscr{T}_{A} ^{T \overline{T} (2)}$};
				\node[] (TBb) [below = of TB]{$\mathscr{T}_{B} ^{\bullet(d>2)} $};
				\draw[<->](TA)--(TB);
				\path[->] (TA) edge [left] node {\text{\footnotesize $T \overline{T}$-deformation}} (TTTA);
				\draw[<->] (TTTA) edge[dashed] (TBb);
				\path[->] (TB) edge[dotted,right] node {\text{\footnotesize deformation?}} (TBb);
			\end{tikzpicture}
	\label{eq:Tdimscheme}
	\end{equation}
As a first step, we can ask the following questions: does the relation
between the theories survive along the RG flow triggered by the $T
\overline{T}$-deformation of \smash{$\mathscr{T}_A^{(2)}$}? If so, what would
be the theory on the other side, denoted \smash{$\mathscr{T}_{B}
^{\bullet(d>2)} $} in the diagram \eqref{eq:Tdimscheme}? Does it admit
a description as a deformation of the theory \smash{$\mathscr{T}_B ^{(d>2)}
$}?

The answers to these questions within a general framework seem out of
reach at the moment. Nevertheless, we hope that the study of the $T
\overline{T}$-deformed $q$-Yang-Mills theory will lead to exciting
discoveries in this direction. Indeed, $q$-Yang-Mills theory is
intimately connected to four-dimensional supersymmetric Yang-Mills
theory in many ways. As described above, one very direct connection is
that it descends from $\mathcal{N}=4$ supersymmetric Yang-Mills theory
on a non-compact four-manifold $X_4$.
There are, however, other tight and direct links between $q$-deformed
two-dimensional Yang-Mills theory and four-dimensional supersymmetric
Yang-Mills theory, for example in the equivalence with the
superconformal index \cite{Gadde:2011ik} in the context of Gaiotto's
$4d$/$2d$ dualities \cite{Gaiotto:2009we}. Two-dimensional
Yang-Mills theory and its deformations appear as well from direct
localization computations on the four-sphere
$\mathbb{S}^4$~\cite{Pestun2dA,PestunGiombiA,PestunGiombiB} and on the
four-dimensional hemisphere~\cite{Wang:2020seq}, possibly with the inclusion of defects \cite{Komatsu:2020sup}.

\paragraph{Acknowledgements.}
We thank Riccardo Conti for fruitful discussions. The work of L.S. was supported by the Doctoral Scholarship
SFRH/BD/129405/2017 from the Funda\c{c}\~{a}o para a Ci\^{e}ncia e a
Tecnologia (FCT). The work of L.S. and M.T. was 
supported by the FCT Project PTDC/MAT-PUR/30234/2017. 
The work of
R.J.S. was supported by the Consolidated Grant ST/P000363/1 from the
UK Science and Technology Facilities Council (STFC). 

\begin{appendix}

\section{Approximate solution for $\boldsymbol{b_{q,\infty}}$}
\label{app:solbqinfty}

In Section \ref{sec:TTphase1} we have studied the weak coupling phase
in the large $N$ limit of $T \overline{T}$-deformed $q$-Yang-Mills
theory. Turning on the $T \overline{T}$-deformation amounts to
replacing $p \mapsto p\, b_{q, \infty}$, where $b_{q, \infty}$ depends
on $p$, $\lambda$ and $\tau$, and is implicitly determined by
\eqref{eq:finftycondition}. In this appendix we analyze
\eqref{eq:finftycondition} in the small $q$-deformation regime, that
is, $p \to \infty$ with $\lambda\, p = A$ fixed. We set $\chi=2$;
the $\chi$-dependence is eventually reinstated by replacing $p$ with
$\frac{2 p}{\chi}$.

\begin{figure}
				\centering
				\includegraphics[width=0.4\textwidth]{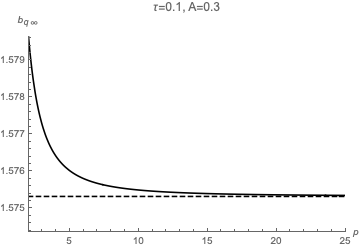}\hspace{0.1\textwidth}
				\includegraphics[width=0.4\textwidth]{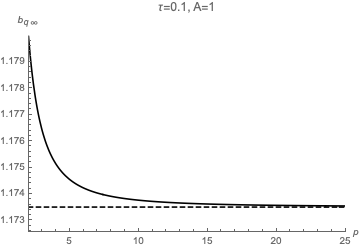}\\[3mm]
				\includegraphics[width=0.4\textwidth]{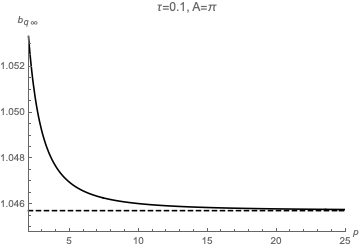}\hspace{0.1\textwidth}
				\includegraphics[width=0.4\textwidth]{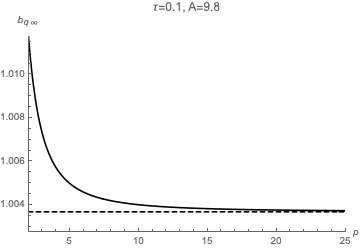}
\caption{\small Plot of $b_{q, \infty}$ as a function of $p$ at
  $\tau=0.1$. The four plots correspond to $A= 0.3, 1,\pi, 9.8$. The
  dashed horizontal line is the corresponding value of $b_{\infty}$ in
  $T \overline{T}$-deformed Yang-Mills theory without
  $q$-deformation.} 
			\label{fig:bq1}
                      \end{figure}

\begin{figure}
				\centering
				\includegraphics[width=0.4\textwidth]{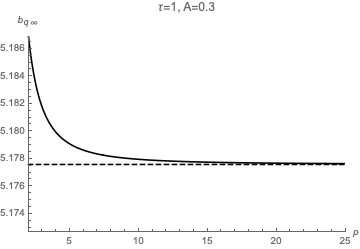}\hspace{0.1\textwidth}
				\includegraphics[width=0.4\textwidth]{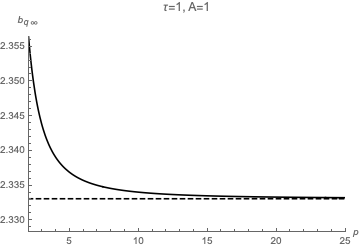}\\[3mm]
				\includegraphics[width=0.4\textwidth]{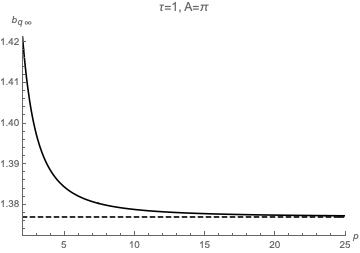}\hspace{0.1\textwidth}
				\includegraphics[width=0.4\textwidth]{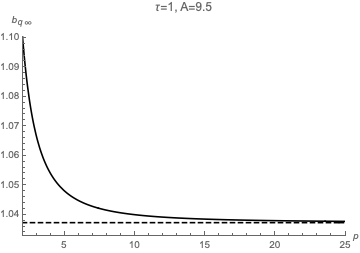}
\caption{\small Plot of $b_{q, \infty}$ as a function of $p$ at
  $\tau=1$. The four plots correspond to $A= 0.3, 1,\pi, 9.5$. The
  dashed horizontal line is the corresponding value of $b_{\infty}$ in
  $T \overline{T}$-deformed Yang-Mills theory without
  $q$-deformation.}
\label{fig:bq2}
\end{figure}

\begin{figure}
				\centering
				\includegraphics[width=0.4\textwidth]{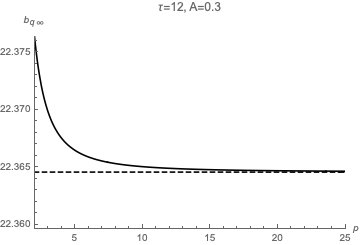}\hspace{0.1\textwidth}
				\includegraphics[width=0.4\textwidth]{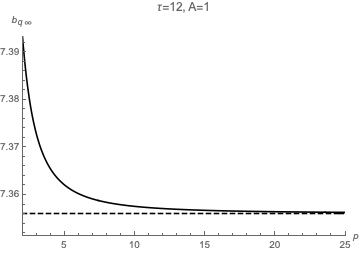}\\[3mm]
				\includegraphics[width=0.4\textwidth]{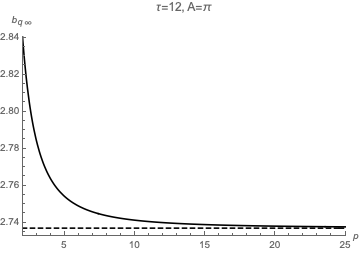}\hspace{0.1\textwidth}
				\includegraphics[width=0.4\textwidth]{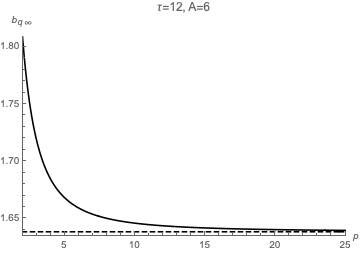}
\caption{\small Plot of $b_{q, \infty}$ as a function of $p$ at
  $\tau=12$. The four plots correspond to $A= 0.3, 1,\pi, 6$. The
  dashed horizontal line is the corresponding value of $b_{\infty}$ in
  $T \overline{T}$-deformed Yang-Mills theory without
  $q$-deformation.} 
\label{fig:bq3}
\end{figure}

Expanding \eqref{eq:finftycondition} at large $p$ we get 
\begin{equation}
\label{eq:approxbinfty}
b_{q, \infty} = \left( 1 - \tau\, \Big( \frac{1}{A\, b_{q, \infty} } +
  \frac{1}{6 p^2\, b_{q, \infty} ^2 } + \frac{A}{72 p^4\, b_{q,
      \infty} ^3} + O\big( p^{-6} \big) - \frac{1}{12} \Big)
\right)^{-2} \ . 
\end{equation}
At $O (p^{-2})$ this equation admits only one solution satisfying
\begin{equation*}
\lim_{\tau\to0}\, b_{q, \infty} = 1 \ ,
\end{equation*}
as required in order to recover the correct behaviour when the $T
\overline{T}$-deformation is turned off. This solution is a decreasing
function of $p$, for fixed $A$ and $\tau$, and converges from above to
the corresponding quantity $b_{\infty}$ of $T \overline{T}$-deformed
Yang-Mills theory as $p$ becomes large. The explicit expression for
this solution is rather lengthy and cumbersome, hence we do not write
it explicitly. Instead, we plot the solution as a function of $p$, for
different values of $A$ and $\tau$, in Figures~\ref{fig:bq1},
\ref{fig:bq2} and~\ref{fig:bq3}.

From these plots we see the qualitative behaviour of $b_{q, \infty}$,
and for all fixed choices of $A$ and $\tau$ it converges
asymptotically to the value $b_{\infty}$. This guarantees that $b_{q,
  \infty}$ indeed arises as a $q$-deformation of $b_{\infty}$, and
$b_{\infty}$ is correctly recovered in the limit
\eqref{eq:qtoYMlimit}. In particular, when $\tau >0$ we have
			\begin{equation*}
				1  < b_{\infty} < b_{q , \infty} \ ,
			\end{equation*}
that is, $b_{q, \infty}$ is a decreasing function of $p$ which
converges to $b_{\infty}$ from above.

\begin{figure}
				\centering
				\includegraphics[width=0.5\textwidth]{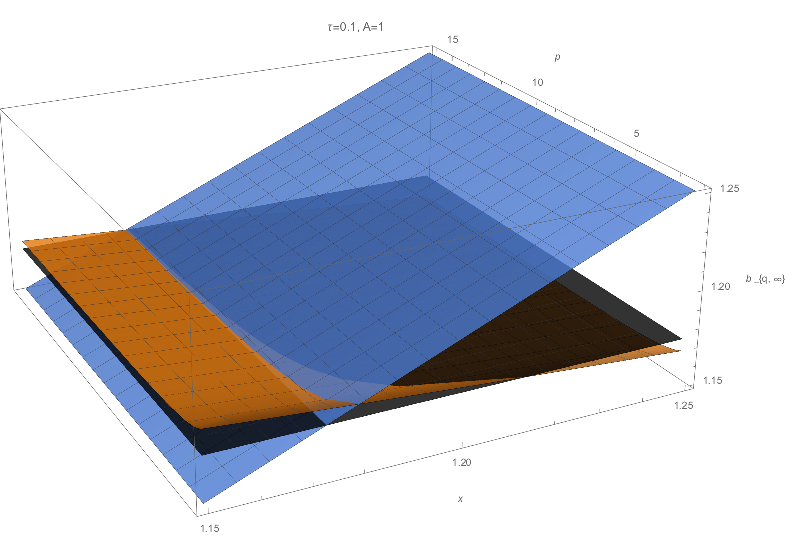}
\caption{\small Plot of $b_{q, \infty}$ as a function of $p$, at $A=1$
  and $\tau=0.1$. The blue surface is the function $x$ on the
  left-hand side of \eqref{eq:approxbinfty}. The orange surface is the
  right-hand side of \eqref{eq:approxbinfty} as a function of $p$ and
  $x$. $b_{q, \infty}$ is determined by the intersection of these two
  surfaces. The black horizontal surface is the asymptotic value~$b_{\infty}$.} 
\label{fig:Op4SolI}
\end{figure}

\begin{figure}
				\centering
				\includegraphics[width=0.5\textwidth]{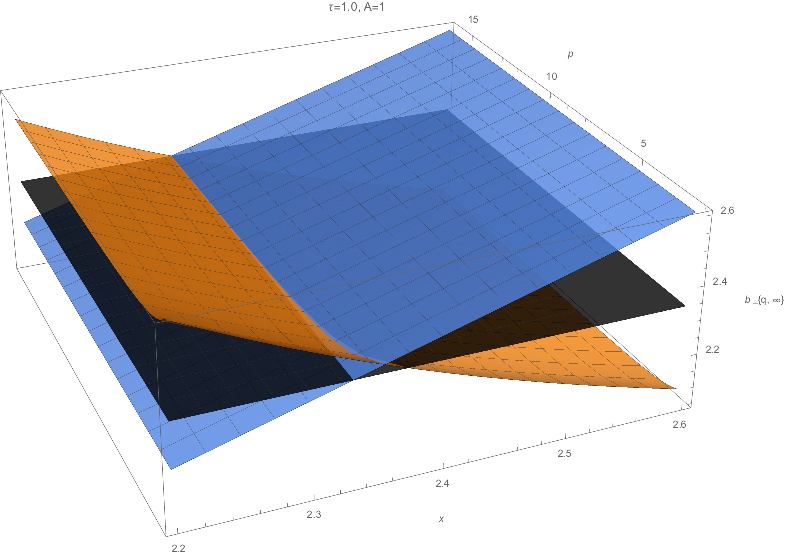}
\caption{\small Plot of $b_{q, \infty}$ as a function of $p$, at $A=1$
  and $\tau=1.0$. The blue surface is the function $x$ on the
  left-hand side of \eqref{eq:approxbinfty}. The orange surface is the
  right-hand side of \eqref{eq:approxbinfty} as a function of $p$ and
  $x$. $b_{q, \infty}$ is determined by the intersection of these two
  surfaces. The black horizontal surface is the asymptotic value~$b_{\infty}$.} 
\label{fig:Op4SolII}
\end{figure}

We now look at how these features are altered when the $O(p^{-4})$
contribution is taken into account. Going to the next non-trivial
order introduces a dependence on $b_{q, \infty} ^{-3}$ in the
right-hand side of \eqref{eq:approxbinfty}. The values of
$b_{q,\infty}$ are thus found by solving a degree seven polynomial
equation, but six solutions will be spurious. We do not dive into an
analytic approach, and instead numerically illustrate the behaviour of
the solution as a function of $p$, for different values of
$\tau$. From what we have learnt at $O(p^{-2})$, it is sufficient to
limit ourselves to $A=1$ and a few values of $\tau$. The solutions are
plotted in Figures \ref{fig:Op4SolI} and \ref{fig:Op4SolII} for $\tau
=0.1$ and $\tau=1$, respectively. We see that the solution is again a
decreasing function of $p$, which changes rapidly for small $p$ and is
almost constant at large $p$. We also check that the solution
approaches $1$ as $\tau \to 0$, in agreement with our analytic
study. The conclusions therefore remain unchanged after the inclusion
of $O(p^{-4})$ corrections. 
			
\section{Instantons in $\boldsymbol{T \overline{T}}$-deformed
  $\boldsymbol q$-Yang-Mills theory}
\label{app:qYMinstantons}
		
In the analyses of \cite{Arsiwalla,Caporaso2005}, where the Poisson
resummation of the heat kernel expansion of the partition function is
done explicitly for the $q$-deformed theory, it was found that the
instantons are responsible for the phase transition also in the
$q$-deformed case. An analogous result was presented in
\cite{JafferisMarsano}, where only the first instanton sector was
taken into account. From the results of~\cite{Santilli2018}, we expect
that this property is not affected by the $T
\overline{T}$-deformation. 
		
\subsubsection*{Two-dimensional Yang-Mills instantons}
		
Let us start with some generalities concerning instantons in
two-dimensional Yang-Mills theory. We start by rewriting the partition
function of $T \overline{T}$-deformed $q$-Yang-Mills theory on $\cs^2$
as in \eqref{eq:Zinstantonexp}: 
\begin{equation}
\label{eq:ZTTinstS2app}
\mz_{\textrm{\tiny $q$-YM}} ^{T \overline{T}} (\lambda,\tau) =
\frac1{N!} \, 
\sum_{\vec\ell \in \Z^N }\, Z_{\vec\ell\,}(\lambda,\tau) \ ,
		\end{equation}
where $Z_{\vec\ell\,}$ encodes the contribution of an instanton labelled by $\vec\ell=(\ell_1,\ell_2,\dots,\ell_N) \in \mathbb{Z}^{N}$. Two-dimensional $U(N)$ Yang-Mills instantons are given by diagonal $\mathfrak{u}(N)$-valued gauge fields 
		\begin{equation*}
			A = \text{diag} \left( A_{\ell_1} ,  A_{\ell_2}  , \dots,  A_{\ell_N} \right)
		\end{equation*}
where $A_{\ell_i}$ is a Dirac monopole potential of charge $\ell_i \in \mathbb{Z}$. Each entry is a gauge connection in the monopole bundle over $\mathbb{S}^2$ of magnetic charge $\ell_i$, 
\begin{equation*}
\mathcal{L}^{ \otimes \ell_i } \longrightarrow \mathbb{S}^2 \ ,
\end{equation*}
where $\mathcal{L}$ is the canonical line bundle of $\mathbb{P}^{1}$
(we identify $\mathbb{P}^{1} \cong \mathbb{S}^2$). Each instanton
configuration determines a splitting of the $U(N)$ gauge bundle on $\cs^2$, which in turn describes a symmetry breaking 
		\begin{equation*}
			U(N) \longrightarrow \prod_{l \in
                          \mathbb{Z}}\, U(N_l) \ , 
		\end{equation*}
where $N_l$ encodes the degeneracy of the magnetic charges: $N_l$
counts how many times the integer $l \in \mathbb{Z}$ appears in the
string $\vec\ell = (\ell_1,\ell_2, \dots, \ell_N ) \in \mathbb{Z}^{N}$
(we omit factors in the product with $N_l=0$). A generic configuration
breaks the gauge group $U(N)$ to its maximal torus $U(1)^N$, while the
one-instanton sector describes a soft breaking $U(N) \to U(1) \times
U(N-1)$. The trivial connection $A=0$ (the zero-instanton sector) is the only gauge field preserving the
full $U(N)$ symmetry.

The restriction to gauge-inequivalent configurations reduces the
coweights $\vec\ell \in \mathbb{Z}^{N}$ to the Weyl chamber 
		\begin{equation*}
			\ell_1 \ge \ell_2 \ge \cdots \ge \ell_N \ ,
		\end{equation*}
but from the symmetry of the partition function we can drop this
restriction at the cost of an overall factor $(N!)^{-1}$. 

\subsubsection*{Complete instanton partition function}

We will now focus on the instanton expansion \eqref{eq:ZTTinstS2app}
and give the first steps towards understanding the complete instanton
partition function, including all instanton contributions. However,
due to the sophistication of the $T \overline{T}$-deformation, we
cannot provide a complete answer and so we only present a partial
analysis here.

Each summand $Z_{\vec\ell\,}$ in \eqref{eq:ZTTinstS2app} can be
computed as the Fourier transform of the contribution $Z_R$ of an
irreducible $U(N)$ representation $R$:
\begin{align*}
Z_{\vec\ell\,}(\lambda,\tau) = \frac{1}{\Delta_q ( \emptyset )^{2}}
\, \int_{\R^N}\, \bigg( \prod_{i=1} ^N\, \dd h_i\, \e^{- 2 \pi \,\I\,
  \ell_i\, h_i } \bigg) \ & \prod_{1 \le i < j \le N}\, 4
  \sinh^2 \frac{ \lambda \left( h_i  - h_j  \right)  }{2N} 
\\ & \times \ \exp \Bigg(  - \frac{
                          \frac{\lambda\, p }{2 N}  \, \Big(
                          \sum\limits_{j=1 } ^{N}\, h_j ^2 - \frac{
                            N\, (N^2 +1)}{12}  \Big) }{ 1 - \frac{
                            \tau}{N^3} \, \Big( \sum\limits_{j=1 }
                          ^{N}\, h_j ^2 - \frac{ N\, (N^2 +1)}{12}
                          \Big) }  \Bigg) \ .
\end{align*}
		We can expand the effect of the $T
                \overline{T}$-deformation in a double power series, finding
		\begin{equation*}
		\begin{aligned}
			Z_{\vec\ell\,}(\lambda,\tau) = K_{N,
                          \lambda,p} \, \int_{\R^N}\,
                        \bigg(\prod_{i=1} ^N & \, \dd h_i\, \e^{ - \frac{
                            \lambda\, p }{2N} \, ( h_i ^2 + \frac{ 4
                            \pi \,\I\, N }{\lambda \, p}\, \ell_i\,
                          h_i )}\bigg) \ \prod_{1 \le i < j \le N}\, 4 \sinh^2 \frac{ \lambda \left( h_i  - h_j  \right) }{2N} \\
				& \times \ \sum_{n=0} ^{\infty}\,
                                \frac{1}{n!} \left( - \frac{ \lambda\,
                                    p }{2 N} \right)^n \ \sum_{k=0}
                                ^{\infty}\, c_{k} (n) \left( \frac{
                                    \tau}{N^3} \right)^k \ \left( \sum_{i=1} ^N\, h_i ^2 - \frac{N\, (N^2 -1)}{12} \right)^{k} \ ,
		\end{aligned}
		\end{equation*}
where $\left\{ c_k (n) \right\}$ are the coefficients of the Taylor
expansion of the function $\big( \frac{x}{1-x} \big)^n$ around $x=0$, and
		\begin{equation*}
			K_{N, \lambda, p } = \e^{\frac{\lambda\, p\, (N^2 - 1)}{24 N}} \, \Delta_q ( \emptyset )^{-2}
		\end{equation*}
		is an overall factor. A direct computation of
                $Z_{\vec\ell\,}$ is difficult already for
                $\tau=0$. However, by exploiting the Weyl denominator
                formula, we can compute a different function
                $\tilde{Z}_{\vec\ell\,}$, which we define in the same
                way as $Z_{\vec\ell\,}$ but with only a single power
                of the $q$-deformed Vandermonde determinant $\Delta_q
                (\vec h)$, instead of the square $\Delta_q (\vec h)^2$
                which enters the expression for $Z_{\vec\ell\,}$. Explicitly, 
\begin{align}
\label{eq:Zelltildeexpansion}
\tilde{Z}_{\vec\ell\,}(\lambda,\tau) = K_{N,
                          \lambda,p} \, \int_{\R^N}\
                        \bigg(\prod_{i=1} ^N & \, \dd h_i\, \e^{ - \frac{
                            \lambda\, p }{2N} \, ( h_i ^2 + \frac{ 4
                            \pi \,\I\, N }{\lambda \, p}\, \ell_i\,
                          h_i )}\bigg) \ \prod_{1 \le i < j \le N}\, 2 \sinh \frac{ \lambda \left( h_i  - h_j  \right) }{2N} \\
				& \times \ \sum_{n=0} ^{\infty}\,
                                \frac{1}{n!} \left( - \frac{ \lambda\,
                                    p }{2 N} \right)^n \ \sum_{k=0}
                                ^{\infty}\, c_{k} (n) \left( \frac{
                                    \tau}{N^3} \right)^k \ \left(
                                  \sum_{i=1} ^N\, h_i ^2 - \frac{N\,
                                    (N^2 -1)}{12} \right)^{k} \ . \nonumber
		\end{align}

At $\tau = 0$, the Weyl denominator formula gives \cite{Arsiwalla,Caporaso2005}
\begin{equation*}
\tilde{Z}_{\vec\ell\,}(\lambda,0) = K^{\prime} _{N, \lambda, p}\, \e^{
  - \frac{ 2 \pi^2\, N }{ \lambda\, p }\, \sum_{i=1} ^N\, \ell_i ^2 } \
\sum_{1 \le i < j \le N}\, \sigma_{ij} \sin \left( \frac{ \ell_i -
    \ell_j }{ 2 p } \right) \ ,
		\end{equation*}
		where $\sigma_{ij}= +1$ if the permutation of the
                first $N$ integers which sends $i$ and $j$ to the
                first and second positions respectively is even, and
                $\sigma_{ij}=-1$ if the permutation is odd. Here
                $K^{\prime} _{N, \lambda, p} $ is another overall
                constant that we will not keep track of.
                
Turning on the $T \overline{T}$-deformation corresponds to introducing powers of the original quadratic Casimir, which in the expansion in \eqref{eq:Zelltildeexpansion} corresponds to introducing the terms with $k>0$. 
The Fourier transform of each summand in \eqref{eq:Zelltildeexpansion}
then gives
		\begin{equation}
		\label{eq:Zelltildesolved}
		\begin{aligned}
			\tilde{Z}_{\vec\ell\,}(\lambda,\tau) &= \tilde{K}_{N,\lambda,p}\, \e^{ - \frac{ 2 \pi^2\, N }{ \lambda\, p }\, \sum_{i=1} ^N\, \ell_i ^2 } \, \sum_{n=0} ^{\infty}\, \frac{1}{n!} \left( - \frac{ \lambda\, p }{2N} \right)^n \ \sum_{k=0} ^{\infty}\, c_{k} (n) \left( \frac{ \tau}{N^3} \right)^k \\
				& \quad \times \ \sum_{1 \le i < j \le
                                  N}\, \sigma_{ij} \left( P^{\rm s}
                                  _{2k} ( \vec\ell\, ) \sin \Big(
                                  \frac{  \ell_i - \ell_j}{ 2 p }
                                  \Big) - \left(\ell_i - \ell_k
                                  \right) P^{\rm c} _{2k} (\vec\ell\,)
                                  \cos \Big( \frac{  \ell_i - \ell_j}{
                                    2 p } \Big) \right) \ ,
		\end{aligned}
		\end{equation}
where $P^{\rm s} _{2k}$ and $P^{\rm c} _{2k}$ are totally symmetric
polynomials of degree $2k$ in $N$ variables, which can be explicitly
computed order by order in $\tau$.

At this point, the expression for $Z_{\vec\ell\,}$ could be obtained
by the Fourier convolution of two functions of the form
\eqref{eq:Zelltildesolved}. The explicit calculation is rather
cumbersome and should be performed order by order in $\tau$; we do
not attempt it here. However, we stress that, for each instanton
sector $\vec\ell$, every order in the perturbative expansion can in
principle be evaluated with the generalization of the strategy
of~\cite{Arsiwalla,Caporaso2005} that we have just sketched.

\end{appendix}

\providecommand{\href}[2]{#2}\begingroup\raggedright\endgroup

\end{document}